\documentclass[12pt]{article}

\usepackage{amstext}
\usepackage{epsfig}

\makeatletter
 
\def\section{\@startsection {section}{1}{\z@}{+3.0ex plus +1ex minus
  +.2ex}{2.3ex plus .2ex}{\normalsize\bf}}
\def\subsection{\@startsection{subsection}{2}{\z@}{+2.5ex plus +1ex
minus +.2ex}{1.5ex plus .2ex}{\normalsize\bf}}
\def\subsubsection{\@startsection{subsubsection}{3}{\z@}{+3.25ex plus
 +1ex minus +.2ex}{1.5ex plus .2ex}{\normalsize\bf}}
 
\expandafter\ifx\csname mathrm\endcsname\relax\def\mathrm#1{{\rm #1}}\fi
 
\makeatletter

\newcount\@tempcntc
\def\@citex[#1]#2{\if@filesw\immediate\write\@auxout{\string\citation{#2}}\fi
  \@tempcnta\z@\@tempcntb\m@ne\def\@citea{}\@cite{\@for\@citeb:=#2\do
    {\@ifundefined
       {b@\@citeb}{\@citeo\@tempcntb\m@ne\@citea
        \def\@citea{,\penalty\@m\ }{\bf ?}\@warning
       {Citation `\@citeb' on page \thepage \space undefined}}%
    {\setbox\z@\hbox{\global\@tempcntc0\csname
b@\@citeb\endcsname\relax}%
     \ifnum\@tempcntc=\z@ \@citeo\@tempcntb\m@ne
       \@citea\def\@citea{,\penalty\@m}
       \hbox{\csname b@\@citeb\endcsname}%
     \else
      \advance\@tempcntb\@ne
      \ifnum\@tempcntb=\@tempcntc
      \else\advance\@tempcntb\m@ne\@citeo
      \@tempcnta\@tempcntc\@tempcntb\@tempcntc\fi\fi}}\@citeo}{#1}}

\def\@citeo{\ifnum\@tempcnta>\@tempcntb\else\@citea
  \def\@citea{,\penalty\@m}%
  \ifnum\@tempcnta=\@tempcntb\the\@tempcnta\else
   {\advance\@tempcnta\@ne\ifnum\@tempcnta=\@tempcntb \else
\def\@citea{--}\fi
    \advance\@tempcnta\m@ne\the\@tempcnta\@citea\the\@tempcntb}\fi\fi}

\def\nl{\nonumber\\}

\newcommand{\lsim}
{\mathrel{\raisebox{-.3em}{$\stackrel{\displaystyle <}{\sim}$}}}
\newcommand{\gsim}
{\mathrel{\raisebox{-.3em}{$\stackrel{\displaystyle >}{\sim}$}}}
\def\asymp#1%
{\mathrel{\raisebox{-.4em}{$\widetilde{\scriptstyle #1}$}}}

\def\Nequal#1%
{\mathrel{\raisebox{-.5em}{$\stackrel{=}{\scriptstyle\rm#1}$}}}

\def\beq#1\eeq{\begin{equation}#1\end{equation}}
\def\beqar{\begin{eqnarray}}
\def\eeqar{\end{eqnarray}}
\def\barr#1{\begin{array}{#1}}
\def\earr{\end{array}}
\def\bfi{\begin{figure}}
\def\efi{\end{figure}}
\def\btab{\begin{table}}
\def\etab{\end{table}}
\def\bce{\begin{center}}
\def\ece{\end{center}}
\def\nn{\nonumber}

\def\arraystretch{1.2}
 
\def\al{\alpha}

\def\Ga{\Gamma}
\def\ga{\gamma}
\def\de{\delta}

\def\eps{\epsilon}
\def\veps{\varepsilon}
\def\la{\lambda}

\def\si{\sigma}
\def\Si{\Sigma}

\def\refeq#1{\mbox{(\ref{#1})}}
\def\refeqs#1{\mbox{(\ref{#1})}}
\def\refeqf#1{\mbox{(\ref{#1})}}
\def\reffi#1{\mbox{Fig.~\ref{#1}}}
\def\reffis#1{\mbox{Figs.~\ref{#1}}}
\def\refta#1{\mbox{Table~\ref{#1}}}

\def\refse#1{\mbox{Section~\ref{#1}}}
\def\refapp#1{\mbox{App.~\ref{#1}}}
\def\citere#1{\mbox{Ref.~\cite{#1}}}
\def\citeres#1{\mbox{Refs.~\cite{#1}}}
 
\newcommand{\TeV}{\unskip\,\mathrm{TeV}}
\newcommand{\GeV}{\unskip\,\mathrm{GeV}}
\newcommand{\MeV}{\unskip\,\mathrm{MeV}}

\newcommand{\ri}{{\mathrm{i}}}

\newcommand{\rd}{{\mathrm{d}}}
\newcommand{\rU}{{\mathrm{U}}}
\newcommand{\rL}{{\mathrm{L}}}
\newcommand{\rT}{{\mathrm{T}}}

\newcommand{\Oa}{\mathswitch{{\cal{O}}(\alpha)}}
\newcommand{\Oaa}{\mathswitch{{\cal{O}}(\alpha^2)}}
\newcommand{\M}{{\cal{M}}}

\def\mathswitchr#1{\relax\ifmmode{\mathrm{#1}}\else$\mathrm{#1}$\fi}
\newcommand{\Pf}{\mathswitch  f}
\newcommand{\Pfbar}{\mathswitch{\bar f}}
\newcommand{\PW}{\mathswitchr W}
\newcommand{\PZ}{\mathswitchr Z}

\newcommand{\PH}{\mathswitchr H}
\newcommand{\Pe}{\mathswitchr e}

\newcommand{\Pd}{\mathswitchr d}
\newcommand{\Pu}{\mathswitchr u}
\newcommand{\Ps}{\mathswitchr s}
\newcommand{\Pc}{\mathswitchr c}
\newcommand{\Pb}{\mathswitchr b}
\newcommand{\Pt}{\mathswitchr t}

\newcommand{\Pep}{\mathswitchr {e^+}}
\newcommand{\Pem}{\mathswitchr {e^-}}
\newcommand{\PWp}{\mathswitchr {W^+}}
\newcommand{\PWm}{\mathswitchr {W^-}}

\newcommand{\PWO}{\mathswitchr {W^3}}
 
\def\mathswitch#1{\relax\ifmmode#1\else$#1$\fi}
\newcommand{\Mf}{\mathswitch {m_\Pf}}

\newcommand{\MW}{\mathswitch {M_\PW}}

\newcommand{\MZ}{\mathswitch {M_\PZ}}
\newcommand{\MH}{\mathswitch {M_\PH}}
\newcommand{\Me}{\mathswitch {m_\Pe}}

\newcommand{\Md}{\mathswitch {m_\Pd}}
\newcommand{\Mu}{\mathswitch {m_\Pu}}
\newcommand{\Ms}{\mathswitch {m_\Ps}}
\newcommand{\Mc}{\mathswitch {m_\Pc}}
\newcommand{\Mb}{\mathswitch {m_\Pb}}
\newcommand{\Mt}{\mathswitch {m_\Pt}}
 
\newcommand{\sw}{\mathswitch {s_\PW}}
\newcommand{\cw}{\mathswitch {c_\PW}}

\newcommand{\GH}{\Ga_\PH}

\def\Re{\mathop{\mathrm{Re}}\nolimits}
\def\ie{i.e.\ }
\def\eg{e.g.\ }
\def\cf{cf.\ }

\newcommand{\cs}{cross-section}
\newcommand{\css}{cross-sections}
\newcommand{\naive}{naive}

\newcommand{\Born}{{\mathrm{Born}}}
\newcommand{\BFM}{{\mathrm{BFM}}}
\newcommand{\pole}{{\mathrm{pole}}}
\newcommand{\onePR}{{\mathrm{1PR}}}

\newcommand{\born}{\mathrm{Born}}
\newcommand{\self}{\mathrm{self}}

\newcommand{\cut}{\mathrm{cut}}

\newcommand{\unpol}{\mathrm{unpol}}

\newcommand{\CMS}{\mathrm{CMS}}

\newcommand{\z}{\setbox0\hbox{+}\hbox to \wd0{\hss0\hss}}
\def\limfunc#1{\mathop{\rm #1}}

\def\Re{\limfunc{Re}}
\def\Im{\limfunc{Im}}
\def\tfrac#1#2{{\textstyle {#1 \over #2}}}
\def\dfrac#1#2{{\displaystyle {#1 \over #2}}}

\def\pslash#1{{\setbox0=\hbox{$#1$}
  \rlap{\ifdim\wd0>.7em\kern.22\wd0\else\kern.1\wd0\fi /}#1}}

\def\braket#1#2{\left\langle #1\vphantom{#2}
  \right. \kern-2.5pt\left| #2\vphantom{#1}\right\rangle }

\def\M{{\cal M}}
\def\O{{\cal O}}
\def\room{\vrule height.7cm depth0pt width0pt}

\def\ZZ{$\PZ\PZ\to\PZ\PZ$}
\def\zzzz{{\PZ\PZ\to\PZ\PZ}}
\def\zzzzL{{\PZ_\rL\PZ_\rL\to\PZ_\rL\PZ_\rL}}
\def\xxxx{{\chi\chi\to\chi\chi}}

\def\SMH{{\cal S}}
\def\TMH{{\cal T}}
\def\UMH{{\cal U}}
\def\OaMH{$\O(\GF\MH^2)$}
\def\OaMH{$\O\left(\al\MH^2/\sw^2\MW^2\right)$}

\def\unpol{\mathrm{unpol}}
\def\tot{\mathswitchr{tot}}
\def\cut{\mathswitchr{cut}}
\def\Pol{\mathswitchr{Polarization}}
\def\entry{\mathswitchr{entry}}
\def\LLLL{\mathswitchr{LLLL}}
\def\LLTT{\mathswitchr{LLTT}}
\def\LTLT{\mathswitchr{LTLT}}
\def\LLLT{\mathswitchr{LLLT}}
\def\LTTT{\mathswitchr{LTTT}}
\def\TTTT{\mathswitchr{TTTT}}
\def\vertex{\mathswitchr{vertex}}
\def\wf{\mathswitchr{wf}}
\def\boxrc{\mathswitchr{box}}
\def\self{\mathswitchr{self}}
\def\reg{\mathswitchr{reg}}
\def\mixed{\mathswitchr{mixed}}
\def\oneloop{\text{1-loop}}
\def\twoloop{\text{2-loop}}

\begin{document}

\thispagestyle{empty}
\def\thefootnote{\fnsymbol{footnote}}
\setcounter{footnote}{1}
\null
\strut\hfill BI-TP 96/37 \\
\strut\hfill PSI-PR-96-31\\
\strut\hfill WUE-ITP-96-023\\
\strut\hfill hep-ph/9612390
\vskip 0cm
\vfill
\begin{center}
{\Large \bf 
\boldmath{Radiative Corrections to $\PZ\PZ\to\PZ\PZ$ \\
    in the Electroweak Standard Model}%
\footnote{Partially supported by the EC network `Human Capital and
Mobility' under contract no.\ CHRX-CT94-0579 and by the 
Bundesministerium f\"ur Bildung, Wissenschaft, Forschung und Technologie
(BMBF) 
under contract no.\ 05 7WZ91P (0).} 
\par} \vskip 2.5em
{\large
{\sc A.~Denner%
}\\[1ex]
{\normalsize \it Paul Scherrer Institut, W\"urenlingen und Villigen\\
CH-5232 Villigen PSI, Switzerland}\\[2ex]
{\sc S.~Dittmaier%
}\\[1ex]
{\normalsize \it Theoretische Physik, Universit\"at Bielefeld\\ 
Postfach 100131, D-33501 Bielefeld, Germany}
\\[2ex]
{\sc T.~Hahn
} \\[1ex]
{\normalsize \it Institut f\"ur Theoretische Physik, Universit\"at W\"urzburg\\
Am Hubland, D-97074 W\"urzburg, Germany}
}
\par \vskip 1em
\end{center} \par
\vskip 1cm 
\vfill
{\bf Abstract:} \par
The \cs\ for $\PZ\PZ\to\PZ\PZ$ with arbitrarily
polarized \PZ\ bosons is calculated within the electroweak
Standard Model including the complete 
${\cal O}(\alpha)$ corrections. 
We show the numerical importance of the radiative corrections and
elaborate its characteristic features.
The treatment of the Higgs-boson resonance is discussed
in different schemes including
the $S$-matrix-motivated pole scheme and the background-field method. 
The numerical accuracy of the equivalence theorem is investigated by
comparing the \css\ for purely longitudinal 
Z bosons
obtained from the equivalence theorem and from the complete calculation.
In this context the full ${\cal O}(\alpha)$ corrections are also
confronted with the enhanced corrections of \OaMH,
which were frequently used in the literature.
\par
\vskip 1cm 
\noindent 
December 1996 \par
\null
\setcounter{page}{0}
\clearpage
\def\thefootnote{\arabic{footnote}}
\setcounter{footnote}{0}

\section{Introduction}

Gauge-boson scattering provides a window into the
heart of spontaneously broken gauge theories: the gauge-boson
self-interactions and the scalar sector, which drives spontaneous
symmetry breaking. Therefore, such processes found continuous
interest in the literature
\cite{Di73,Le77,Pa85,Da89,Ma89,ETVVVV,Gu93,VVVVcorr}
since the very first years of spontaneously broken gauge theories.
Since lowest-order predictions for all gauge-boson scattering
amplitudes involve only interactions between gauge and
scalar bosons, the corresponding \css\ depend very sensitively
on the non-abelian and scalar sector of the underlying theory. 
This sensitivity is even enhanced for high-energetic, longitudinally 
polarized massive gauge bosons, 
owing to the presence of gauge cancellations.
A longitudinal polarization vector 
contains a factor $k^0/M$, where $k^\mu$ and $M$ are the momentum and
mass of the corresponding gauge boson, respectively,
and induces contributions to the matrix element that grow with energy.
In spontaneously broken gauge theories such 
contributions cancel in the high-energy limit,
as required by unitarity.
For `t~Hooft gauge-fixing conditions these so-called
unitarity cancellations are quantitatively expressed by the 
Goldstone-boson equivalence theorem (ET) \cite{et,Ch85,Ya88,Gr95,bgfet} 
which relates amplitudes for longitudinal gauge bosons to 
those of the corresponding Goldstone bosons and thus reflects
the connection between gauge and scalar sector of the theory.

In the minimal $\mathrm{SU}(2)\times \mathrm{U}(1)$ 
electroweak Standard Model (SM) only
one physical scalar field remains after spontaneous symmetry breaking, 
viz.\ the Higgs boson, which plays a central role in the discussion of
massive gauge-boson scattering. 
Virtual Higgs-boson exchange is needed to prevent the 
$2\to2$ scattering amplitudes of longitudinal gauge bosons from violating 
the (perturbative) unitarity bound at high energies. 
In turn, the requirement of unitarity can be used to derive bounds on
the Higgs-boson mass below
which the SM remains weakly interacting and
treatable in low-order perturbation theory \cite{Di73,Le77}.
These bounds are of the order of 1 TeV and are slightly strengthened 
by including the \OaMH\ radiative corrections (RCs) to gauge-boson scattering
\cite{Da89}.
As already pointed out in \citere{Pa85},
all these bounds are
only qualitative, 
since they are obtained by applying perturbation theory
in a region where it breaks down.
The bounds on $\MH$ can be related to a scale of new  physics,
which is necessary to avoid the Landau pole in the scalar self-interaction
\cite{Ma89}.
If these bounds are not satisfied 
the Higgs sector becomes strongly interacting. In this case 
large effects of new physics
should arise in gauge-boson scattering and these processes 
would be particularly suited to study the
electroweak symmetry breaking sector of the SM \cite{Le77,Ch85,Ve77}.

Gauge-boson scattering reactions can be studied at all high-energy
colliders, \ie pp colliders like the  LHC, $\Pep\Pem$ colliders like the NLC, 
or $\mu^+\mu^-$ colliders, where 
these reactions 
naturally appear as subprocesses. 
At high energies ($E\gg\MW$) the 
incoming particles radiate plenty of gauge bosons.  
Similar to the well-known Weizs\"acker--Williams approximation for
photonic reactions also massive vector-boson scattering at high energies
can be approximated by convoluting the vector-boson cross-section with
the corresponding flux of gauge bosons. This approximation is known as
{\it equivalent vector-boson method\/} (see e.g.\ \citere{Ku96} and
references therein).

At high energies,
where the investigation of gauge-boson scattering
is most interesting, the RCs are typically large and need to be taken into
account.
In this paper we investigate the effects of 
RCs on on-shell massive gauge-boson scattering processes. 
We have chosen the simplest representative,
the process $\PZ\PZ\to\PZ\PZ$. It contains all interesting features 
that are typical for massive gauge-boson scattering such as
the occurrence of a Higgs-boson resonance or enhanced 
RCs associated with a heavy Higgs boson. On the other hand,
complications by bremsstrahlung corrections, which occur for W bosons,
are absent.

We calculate the complete ${\cal O}(\alpha)$ RCs to 
$\PZ\PZ\to\PZ\PZ$ and present a detailed numerical discussion of the 
${\cal O}(\alpha)$-corrected 
\css\ both for the unpolarized case
and the most interesting individual polarizations. 
Once RCs are taken into account,
the introduction of a finite decay width of the Higgs boson, which is
necessary for a sensible description of the resonance, is non-trivial 
owing to problems with gauge invariance.
We compare different treatments such as the \naive\
introduction of a finite width, 
Laurent expansions about the complex pole,
as well as Dyson summation of self-energy corrections. 
The latter procedure is, in particular, 
applied within the framework
of the background-field method (BFM)
(see \citere{bgflong} and references therein), 
where Dyson summation does
not disturb the underlying Ward identities \cite{bgfet} which guarantee 
gauge cancellations and unitarity. 

For longitudinal gauge-boson scattering the radiative corrections of
\OaMH, which dominate for a heavy Higgs boson, have been calculated in
the literature using the ET. We test the accuracy of such an approach by
comparing these results with the full $\Oa$ corrections. Moreover, we
have calculated the $\Oa$ corrections as predicted via the ET, which
possess a very simple analytical form.

This paper is organized as follows:
After some preliminary remarks in \refse{se:prelim} about kinematics,
conventions, and discrete symmetries, we discuss the lowest-order
cross-sections in \refse{se:locs}. In \refse{se:rcs} we describe the
explicit calculation and the structure of the $\Oa$ corrections. The
different methods for introducing a finite Higgs-boson width are
presented in \refse{se:Hres}. A brief description of the application of
the ET to $\zzzzL$ and the 
heavy-Higgs-boson effects in \refse{se:zzzzL}
concludes our presentation of the calculational framework. Numerical
results are discussed in \refse{se:numres}, 
and Section~\ref{se:concl}
contains our conclusions. 
Appendix~\ref{app:d0sing} provides 
a further discussion of the Landau singularity that appears in 
some box diagrams.
In Appendix~\ref{se:ETexpl} we present the full analytical results for
the ${\cal O}(\alpha)$ corrections obtained via the ET.

\section{Preliminaries}
\label{se:prelim}

\subsection{Kinematics and conventions}

We consider the reaction
\begin{equation}
Z(k_1,\la _1)+Z(k_2,\la _2)\to Z(k_3,\la _3)+Z(k_4,\la _4)\,,
\end{equation}
where $k_i$ and $\la_i$ denote the momenta and helicities of the
incoming and outgoing \PZ~bosons,
respectively.
We use the indices $\rL$, $\rT$, and $\rU$ to indicate longitudinal ($\la=0$),
transverse 
($\la = \pm$), and unpolarized \PZ~bosons, respectively,
and characterize definite polarization combinations by a sequence of
four letters, \eg LTLT stands for $\PZ_\rL\PZ_\rT\to\PZ_\rL\PZ_\rT$.

The incoming particles travel along the $z$ axis and are scattered into
the $x$--$z$ plane. In the center-of-mass system (CMS) 
the momenta and polarization vectors $\veps_i(\la_i)$ read 
\begin{equation}
\begin{array}[b]{rllrl}
\label{polvecs}
\vphantom{\tfrac 1{\sqrt 2}}
k_1^\mu = & \left( E,\,0,\,0,\,-p \right), & &
k_3^\mu = & \left( E,\,-p\sin \theta ,\,0,\,-p \cos \theta \right), \\
\varepsilon _1^\mu (0) = & \left( -p,\,0,\,0,\,E \right)/\MZ, & &
\varepsilon _3^{\mu ,\ast } (0) = & 
	\left( p,\,-E \sin \theta ,\,0,\,-E\cos \theta \right)/\MZ, \\
\varepsilon _1^\mu (\pm ) = & \left( 0,\,-1,\,\pm \ri,\,0 \right)/\sqrt{2}, & &
\varepsilon _3^{\mu ,\ast }(\pm ) = & 
	\left( 0,\,-\cos \theta ,\,\mp \ri,\,\sin \theta \right)/\sqrt{2}, \\ 
\room
\vphantom{\tfrac 1{\sqrt 2}}
k_2^\mu = & \left( E,\,0,\,0,\,p \right), & &
k_4^\mu = & \left( E,\,p \sin \theta ,\,0,\,p \cos \theta \right), \\
\varepsilon _2^\mu (0) = & \left( -p,\,0,\,0,\,-E \right)/\MZ, \quad & &
\varepsilon _4^{\mu ,\ast }(0) = & 
	\left( p,\,E\sin \theta ,\,0,\,E\cos \theta \right)/\MZ, \\
\varepsilon _2^\mu (\pm ) = & \left( 0,\,1,\,\pm \ri,\,0 \right)/\sqrt{2}, & &
\varepsilon _4^{\mu ,\ast }(\pm ) = & 
	\left( 0,\,\cos \theta ,\,\mp \ri,\,-\sin \theta \right)/\sqrt{2}
\end{array}
\end{equation}
in terms of the energy $E$ of the \PZ~bosons, their momentum 
$p=\sqrt{E^2-\MZ^2}$, and the scattering angle $\theta$.
The Mandelstam variables are defined as
\begin{equation}
\begin{array}[b]{rclcl}
s & = & (k_1+k_2)^2 & = & 4E^2\,, \\
t & = & (k_1-k_3)^2 & = & -4p^2\sin ^2\theta /2\,, \\
u & = & (k_1-k_4)^2 & = & -4p^2\cos ^2\theta /2\,.
\end{array}
\end{equation}

Following the treatment of \citere{aaww} for $\ga\ga\to\PW\PW$, we
introduce the 83 standard matrix elements (SMEs) $\M_{ijkl}$ which
contain the complete
information about the boson polarizations%
\footnote{Only 81 SMEs are linearly independent, the other two are kept
for convenience.}.
The invariant matrix element $\M$ is decomposed 
into a linear combination of
the SMEs with invariant functions $F_{ijkl}(s,t)$ as coefficients.
Exploiting discrete symmetries, the number of independent SMEs can be
reduced.

In terms of the invariant matrix element $\M$
the differential \cs\ is expressed as
\begin{equation}
\left(\frac{\rd\si}{\rd\Omega}\right)_{\la_1\la_2\la_3\la_4}
=\frac 1{64\pi^2 s}
\left| \M_{\la_1\la_2\la_3\la_4} \right| ^2\,.
\end{equation}
The unpolarized \cs\ results from an average over the 
initial states and a sum over the final states,
\begin{equation}
\left( \frac{\rd\sigma }{\rd\Omega }\right) _{\unpol}
=\frac 19\sum _{\lambda _1,\lambda_2} \,\sum _{\lambda _3,\lambda _4}
\left( \frac{\rd\sigma }{\rd\Omega }\right) 
_{\lambda_1\lambda_2\lambda_3\lambda_4}\,.
\end{equation}
More generally, the correct
average is obtained by multiplying with 
$1/3$ for each unpolarized \PZ~boson and by $1/2$ for each
transverse \PZ~boson in the initial state.

The integrated \cs\ is obtained by 
\begin{equation}
\si_{\tot}=\frac 12\,\int _{0^\circ}^{360^\circ} \rd\varphi \int _{\theta 
_{\cut}} ^{180^\circ-\theta _{\cut}} \rd\theta \,\sin 
\theta \,\frac{\rd\sigma }{\rd\Omega }\,,
\end{equation}
where $\theta _{\cut}$ denotes an angular cut
which is set to $10^\circ$ 
in our numerical evaluations.
The symmetry factor $1/2$ results from the presence of
two identical particles in the final state.

\subsection{Discrete symmetries}

As a consequence of Bose symmetry the amplitude $\M$ is invariant under
the interchange $(k_1,\varepsilon_1) \leftrightarrow (k_2,\varepsilon_2)$ 
and/or $(k_3,\varepsilon^*_3) \leftrightarrow (k_4,\varepsilon^*_4)$, \ie
\beqar
{\cal M}_{\lambda_1 \lambda_2 \lambda_3 \lambda_4}(E,\theta) &=&
{\cal M}_{\lambda_2 \lambda_1 \lambda_4 \lambda_3}(E,\theta) \nl
&=& {\cal M}_{\lambda_1 \lambda_2 \lambda_4 \lambda_3}(E,180^\circ+\theta) 
= {\cal M}_{\lambda_2 \lambda_1 \lambda_3 \lambda_4}(E,180^\circ+\theta).
\eeqar
This implies for the \css
\beqar \label{boserel}
\left(\frac{\rd\sigma}{\rd\Omega}\right)_{\la_1\la_2\la_3\la_4}(s,t) &=&
\left(\frac{\rd\sigma}{\rd\Omega}\right)_{\la_2\la_1\la_4\la_3}(s,t) \nl
&=&\left(\frac{\rd\sigma}{\rd\Omega}\right)_{\la_2\la_1\la_3\la_4}(s,u)=
\left(\frac{\rd\sigma}{\rd\Omega}\right)_{\la_1\la_2\la_4\la_3}(s,u) .
\eeqar
In particular, all \css\ with equally polarized incoming and/or
outgoing \PZ~bosons are forward--backward symmetric.

CPT symmetry entails 
\beq \label{CPTrel}
{\cal M}_{\lambda_1 \lambda_2 \lambda_3 \lambda_4} =
{\cal M}_{\lambda_3 \lambda_4 \lambda_1 \lambda_2}
\eeq
and the analogous relation for the \css.

Because quark mixing is completely negligible for this process,
we use a unit quark-mixing matrix, and thus
also CP is an exact symmetry%
\footnote{Even for a general quark-mixing matrix, CP would be violated
in the considered process only beyond the one-loop level.}.
As a consequence, the helicity amplitudes are related as follows
\beq
{\cal M}_{\lambda_1 \lambda_2 \lambda_3 \lambda_4} =
{\cal M}_{-\lambda_1 -\lambda_2 -\lambda_3 -\lambda_4},
\eeq
and the \css\ do not change if all helicities are reversed.

Owing to CP invariance all SMEs involving the totally antisymmetric Levi-Civita
tensor drop out, and only 43 SMEs 
can appear.
As a consequence of Bose and CPT symmetry only the sum of each SME 
and the ones obtained from the interchanges
$(\varepsilon_1,k_1,\varepsilon^*_3,k_3)\leftrightarrow
 (\varepsilon_2,k_2,\varepsilon^*_4,k_4)$ 
and 
$(\varepsilon_1,k_1,\varepsilon_2,k_2)\leftrightarrow
 (\varepsilon^*_3,k_3,\varepsilon^*_4,k_4)$
occur.
This leaves 19 SMEs, among which 17 are independent.

\section{Lowest-order \cs}
\label{se:locs}

To lowest order, only the three diagrams in Fig.\ \ref{borndiags} 
contribute and yield the following amplitude:
\begin{figure}
\begin{center}
\epsfig{figure=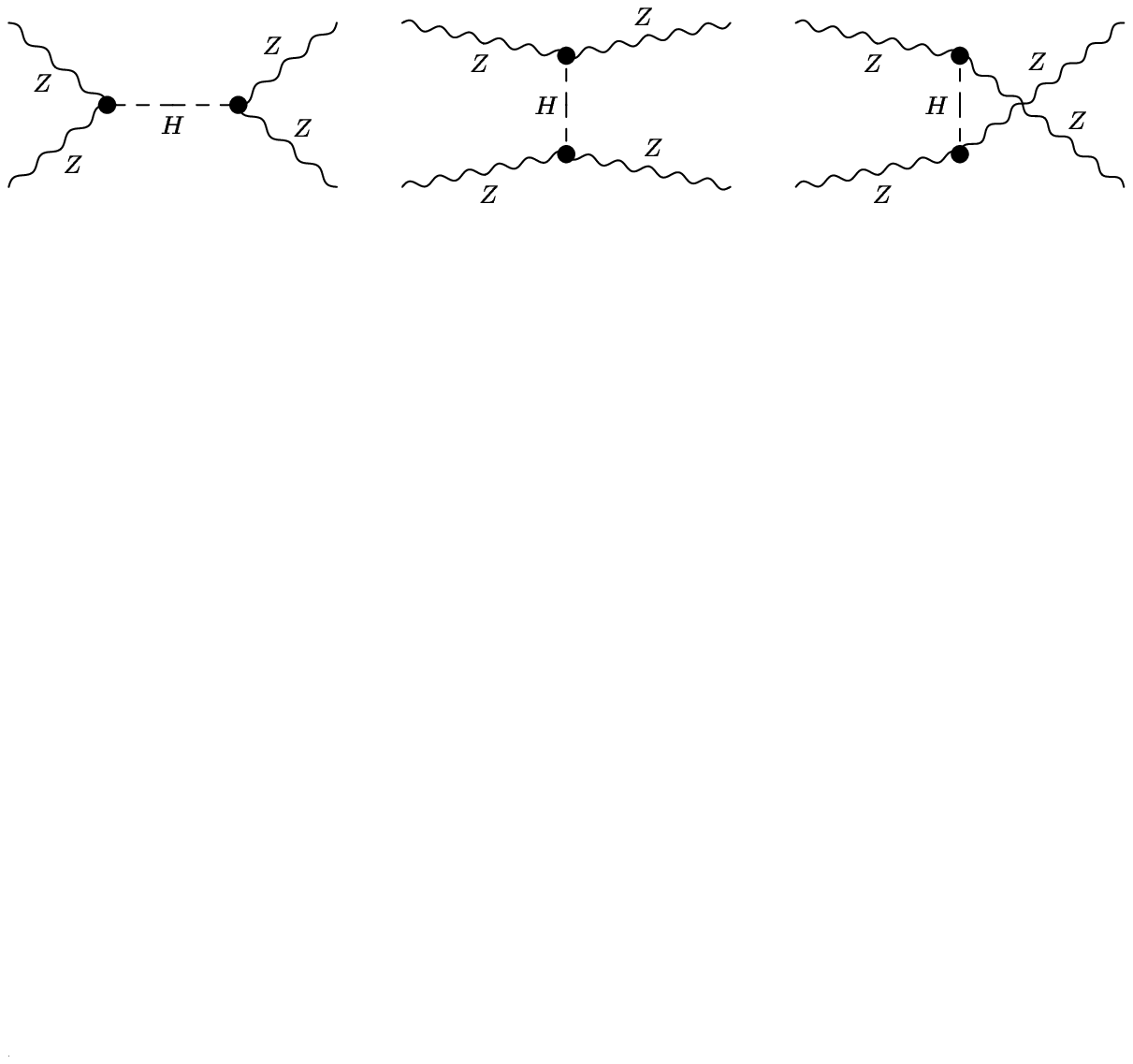}
\end{center}
\caption{\label{borndiags}Lowest-order diagrams for \ZZ }
\end{figure}%
\begin{equation}
\label{bornamp}
\M_{\Born}=-\frac{e^2\MZ^2}{\cw^2\sw^2}\left(
\frac{\M _{00}^{(s)}}{s-\MH^2}+
\frac{\M _{00}^{(t)}}{t-\MH^2}+
\frac{\M _{00}^{(u)}}{u-\MH^2}\right) \,,
\end{equation}
where $\cw=\cos\theta_\PW=\MW/\MZ$, $\sw=\sin\theta_\PW=\sqrt{1-\cw^2}$, 
and the relevant 
SMEs $\M_{00}^{(r)}$ are given by
\begin{equation}  
\M _{00}^{(s)} = (\varepsilon _1\cdot \varepsilon _2 )
(\varepsilon _3^\ast \cdot \varepsilon _4^\ast ) \,,\qquad
\M_ {00}^{(t)}  =  (\varepsilon _1 \cdot \varepsilon _3^\ast )
(\varepsilon _2\cdot \varepsilon _4^\ast ) \,,\qquad
\M _{00}^{(u)}  =  (\varepsilon _1 \cdot \varepsilon _4^\ast )
(\varepsilon _2\cdot \varepsilon _3^\ast ) \,.
\label{eq:M00}
\end{equation}
Explicit formulae for $\M_\born$ for the 81 polarization 
combinations are listed in \refta{ta:bornme}.
\begin{table}
\renewcommand\arraystretch{1.0}
\def\hite{\vrule width0pt depth3.6ex height4.6ex}
\def\rL{\rm L}
{\footnotesize
$$
\begin{array}{|c|c||c|c|} \hline
\vrule depth2ex height3.5ex width0pt
\ \Pol\ &
\M_\Born =-\tfrac{e^2}{\cw^2\sw^2}\times (\entry)
&
\ \Pol\ &
\M_\Born =-\tfrac{e^2}{\cw^2\sw^2}\times (\entry)
\\ \hline \hline
\begin{array}{l}
\rL\rL\rL\rL
\end{array} &
\multicolumn{3}{|c|}{\hite
\dfrac 1{4\MZ^2}\left[
\dfrac{(s-2\MZ^2)^2}{\SMH }+\dfrac{(s(u-t)+\tilde s^2)^2}{4
\tilde s^2\,\TMH }+
\dfrac{(s(t-u)+\tilde s^2)^2}{4\tilde s^2\,\UMH }\right] }
\\ \hline
\begin{array}{l}
{\pm}\rL\rL\rL \\
\rL{\pm}\rL\rL \\
\rL\rL{\pm}\rL \\
\rL\rL\rL{\pm}
\end{array} &
\multicolumn{3}{|c|}{\hite
\dfrac 1{4\sqrt 2\MZ}\dfrac{\sqrt{stu}}{\tilde s^2}\left[
\dfrac{s(u-t)+\tilde s^2}{\TMH }-\dfrac{s(t-u)+\tilde s^2}{\UMH }\right] }
\\ \hline
\begin{array}{l}
\rL\rL{\pm}{\pm} \\
{\pm}{\pm}\rL\rL
\end{array} &
\hite
\dfrac{s-2\MZ^2}{2\SMH } + \dfrac{stu}{2\tilde s^2}\left[
\dfrac 1{\TMH }+\dfrac 1{\UMH }\right]
&
\begin{array}{l}
\rL\rL{\pm}{\mp} \\
{\pm}{\mp}\rL\rL
\end{array} & 
\hite
\dfrac{stu}{2\tilde s^2}\left[ \dfrac 1{\TMH }+\dfrac 1{\UMH }\right]
\\ \hline
\begin{array}{l}
{\pm}\rL{\pm}\rL \\
\rL{\pm}\rL{\pm}
\end{array} &
\ \hite
\dfrac u{4\tilde s^2}\left[-\dfrac{s(u-t)+\tilde s^2}{\TMH }
+\dfrac{2st}{\UMH }\right] \ 
&
\begin{array}{l}
{\pm}\rL{\mp}\rL \\
\rL{\pm}\rL{\mp}
\end{array} &
\ \hite
\dfrac t{4\tilde s^2}\left[\dfrac{s(u-t)+\tilde s^2}{\TMH }+
\dfrac{2su}{\UMH }\right] \
\\ \hline
\begin{array}{l}
{\pm}\rL\rL{\pm} \\
\rL{\pm}{\pm}\rL
\end{array} &
\hite
\dfrac t{4\tilde s^2}\left[\dfrac{2su}{\TMH }-
\dfrac{s(t-u)+\tilde s^2}{\UMH }\right]
&
\begin{array}{l}
{\pm}\rL\rL{\mp} \\
\rL{\pm}{\mp}\rL
\end{array} &
\hite
\dfrac u{4\tilde s^2}\left[ \dfrac{2st}{\TMH }+
\dfrac{s(t-u)+\tilde s^2}{\UMH }\right]
\\ \hline
\begin{array}{l}
\rL{\pm}{\pm}{\pm} \\
{\pm}\rL{\pm}{\pm} \\
{\pm}{\pm}\rL{\pm} \\
{\pm}{\pm}{\pm}\rL
\end{array} & 
\hite
-\dfrac{\MZ}{\sqrt 2}\dfrac{\sqrt{stu}}{\tilde s^2}\left[
\dfrac u{\TMH }-\dfrac t{\UMH }\right]
&
\begin{array}{l}
\rL{\mp}{\pm}{\pm} \\
{\mp}\rL{\pm}{\pm} \\
{\pm}{\pm}\rL{\mp} \\
{\pm}{\pm}{\mp}\rL
\end{array} &
\hite
\dfrac{\MZ}{\sqrt 2}\dfrac{\sqrt{stu}}{\tilde s^2}\left[
\dfrac t{\TMH }-\dfrac u{\UMH }\right]
\\ \hline
\begin{array}{l}
\rL{\pm}{\mp}{\pm} \\
{\pm}\rL{\pm}{\mp} \\
{\mp}{\pm}\rL{\pm} \\
{\pm}{\mp}{\pm}\rL
\end{array} &
\hite
-\dfrac{\MZ}{\sqrt 2}\dfrac{u\sqrt{stu}}{\tilde s^2}\left[
\dfrac 1{\TMH }+\dfrac 1{\UMH }\right]
&
\begin{array}{l}
\rL{\pm}{\pm}{\mp} \\
{\pm}\rL{\mp}{\pm} \\
{\pm}{\mp}\rL{\pm} \\
{\mp}{\pm}{\pm}\rL
\end{array} & 
\hite
\dfrac{\MZ}{\sqrt 2}\dfrac{t\sqrt{stu}}{\tilde s^2}\left[
\dfrac 1{\TMH }+\dfrac 1{\UMH }\right]
\\ \hline
\begin{array}{l}
{\pm}{\pm}{\pm}{\mp} \\
{\pm}{\pm}{\mp}{\pm} \\
{\pm}{\mp}{\pm}{\pm} \\
{\pm}{\mp}{\mp}{\mp}
\end{array} & 
-\MZ^2\dfrac{tu}{\tilde s^2}\left[ \dfrac 1{\TMH }+\dfrac 1{\UMH }\right]
& \multicolumn{2}{|l|}{}
\\ \hline
\begin{array}{l}
{\pm}{\mp}{\pm}{\mp}
\end{array} & 
\hite
\MZ^2\dfrac{u^2}{\tilde s^2}\left[
\dfrac 1{\TMH }+\dfrac 1{\UMH }\right]
&
\begin{array}{l}
{\pm}{\mp}{\mp}{\pm}
\end{array} & 
\hite
\MZ^2\dfrac{t^2}{\tilde s^2}\left[
\dfrac 1{\TMH }+\dfrac 1{\UMH }\right]
\\ \hline
\begin{array}{l}
{\pm}{\pm}{\pm}{\pm}
\end{array} & 
\hite
\MZ^2\left[
\dfrac 1{\SMH }+\dfrac{u^2}{\tilde s^2\TMH }+\dfrac{t^2}
{\tilde s^2\UMH }\right]
&
\begin{array}{l}
{\pm}{\pm}{\mp}{\mp}
\end{array} & 
\hite
\MZ^2\left[
\dfrac 1{\SMH }+\dfrac{t^2}{\tilde s^2\TMH }+\dfrac{u^2}
{\tilde s^2\UMH }\right]
\\ \hline
\end{array}
$$}
\caption{Polarized lowest-order matrix elements 
($\SMH =s-\MH^2$, $\TMH =t-\MH^2$, $\UMH =u-\MH^2$, and $\tilde s=s-4\MZ^2$)}
\label{ta:bornme}
\end{table}

The dimensionful $ZZH$ coupling leads to a suppression of the Born matrix 
element by a factor $\MZ^2/|s-\MH^2|$ for $|s-\MH^2|\gg\MZ^2$.
As a consequence, the lowest-order matrix element for purely transverse 
\PZ~bosons is suppressed for high energies, $s\gg\MH^2$, by a factor $\MZ^2/s$.
Each longitudinal \PZ~boson introduces a factor $\sqrt{s}/\MZ$ 
via its polarization vector.
For helicity amplitudes with more than two external
longitudinal \PZ~bosons unitarity cancellations take place such that 
the Born matrix element with purely longitudinal \PZ~bosons approaches
a constant and
those with three longitudinal \PZ~bosons behave as $\MZ/\sqrt{s}$ at
high energies.
As a remnant of the unitarity cancellations the matrix elements
involving four and three longitudinal \PZ~bosons 
are enhanced by a factor $\MH^2/\MZ^2$ 
if $\MZ^2\ll\MH^2\ll s$.

The analytical 
results for the asymptotic behavior of the
integrated \css\ 
at high energies ($s\gg\MZ^2,\MH^2$)
are listed in \refta{ta:highenergy}.
\begin{table}
\def\bigheight{\vrule width0pt height4.5ex depth3.5ex}
\def\cc{\cos\theta_\cut}
\def\scq{\sin^2\theta_\cut}
$$
\setbox0\hbox{\Pol}
\begin{array}{|c|c||c|c||c|c|} \hline
\bigheight
\Pol & 
\multicolumn{5}{|c|}{
\si_\Born= 
2\pi\,\dfrac{1}{64\pi^2s}\,\dfrac{e^4}{\cw^4\sw^4}\times(\entry) 
}
\\ \hline\hline
\bigheight
\hbox to \wd0{\hfill\LLLL\hfill} & \hfill
\hbox to 1.6\wd0{\hfill$\dfrac{9\MH^4}{16\MZ^4}\cc$\hfill} &
\hbox to \wd0{\hfill\LLTT\hfill} &
\hbox to 1.2\wd0{\hfill$ \dfrac{\cc}{2}$\hfill} &
\hbox to \wd0{\hfill\LTLT\hfill} &
\hbox to 1.1\wd0{\hfill$ \dfrac{\cc}{4}$\hfill}
\\ \hline
\bigheight
\LLLT &
\multicolumn{5}{|c|}{
    \dfrac{(\MH^2+2\MZ^2)^2}{2\MZ^2s}
    \left\{ \ln \dfrac{s\,(1+\cc)}{2\MH^2+s(1-\cc)}-
    \dfrac{2\cc\,(6\MH^2+s\scq)}
      {4\MH^2+s\scq} \right\}
}
  \\ \hline
\bigheight
\LTTT &
\multicolumn{5}{|c|}{
    \dfrac{\MZ^2}{s}
    \left\{ 2\ln \dfrac{s\,(1+\cc)}{2\MH^2+s(1-\cc)}-
    \dfrac{\cc\,(20\MH^2+3s\scq)}
      {4\MH^2+s\scq} \right\}
}
  \\ \hline
\bigheight
\TTTT &
\multicolumn{5}{|c|}{
    \dfrac{\MZ^4}{s^2}
    \left\{ -8\ln\dfrac{s\,(1+\cc)}{2\MH^2+s(1-\cc)}
    +\dfrac{\cc\, s(8+11\scq)}{4\MH^2+s\scq} \right\}
}
  \\ \hline
\end{array}
$$
\caption{Polarized lowest-order \css\ at high energies ($s\gg\MZ^2,\MH^2$)
integrated over $\theta_\cut<\theta<180^\circ-\theta_\cut$}
\label{ta:highenergy}
\end{table}%
\begin{figure}
\epsfig{figure=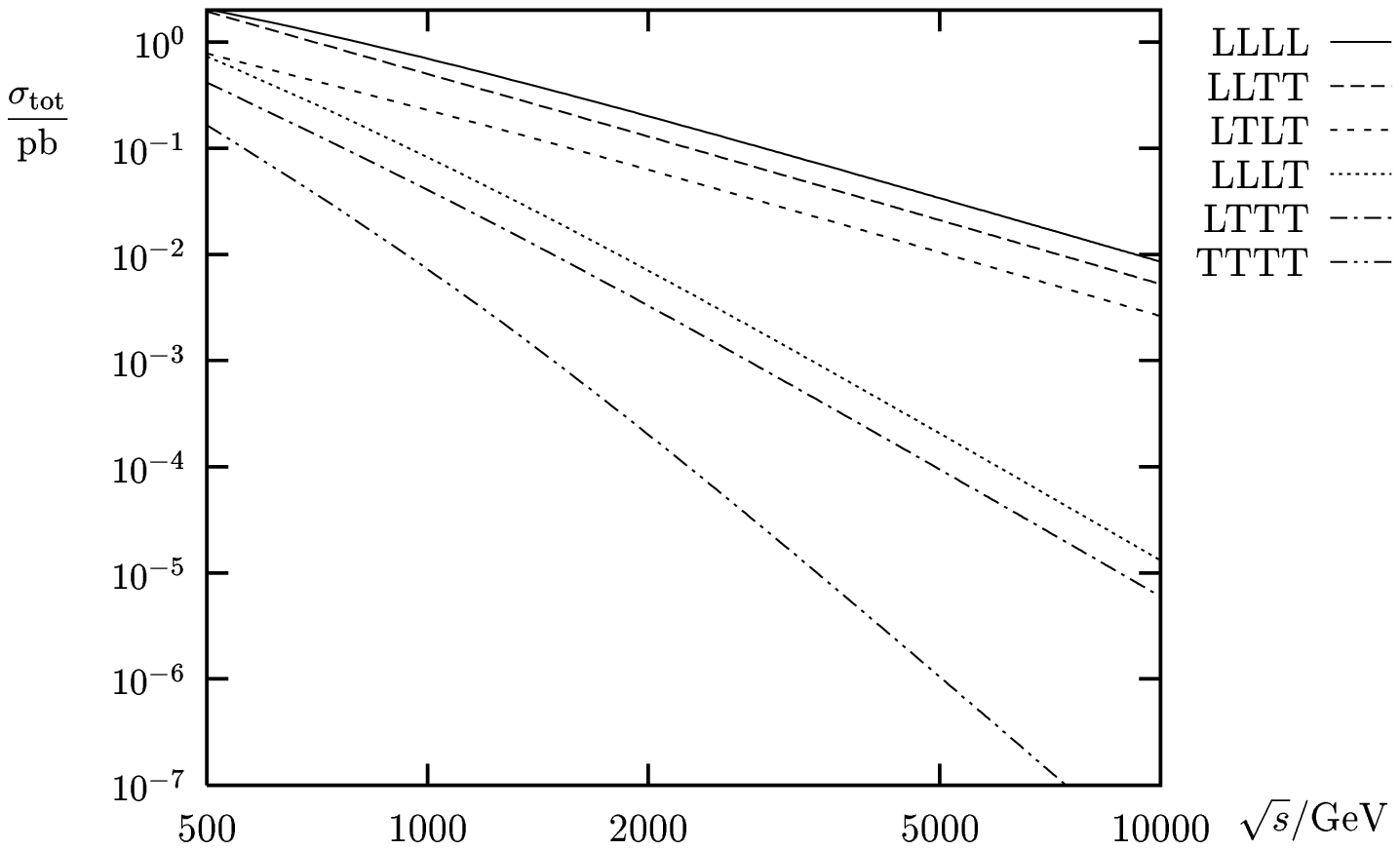}
\caption{Integrated lowest-order \css\ for various polarizations
at $\MH=100\GeV$}
\label{fi:born.MH=100}
\vspace{2em}
\epsfig{figure=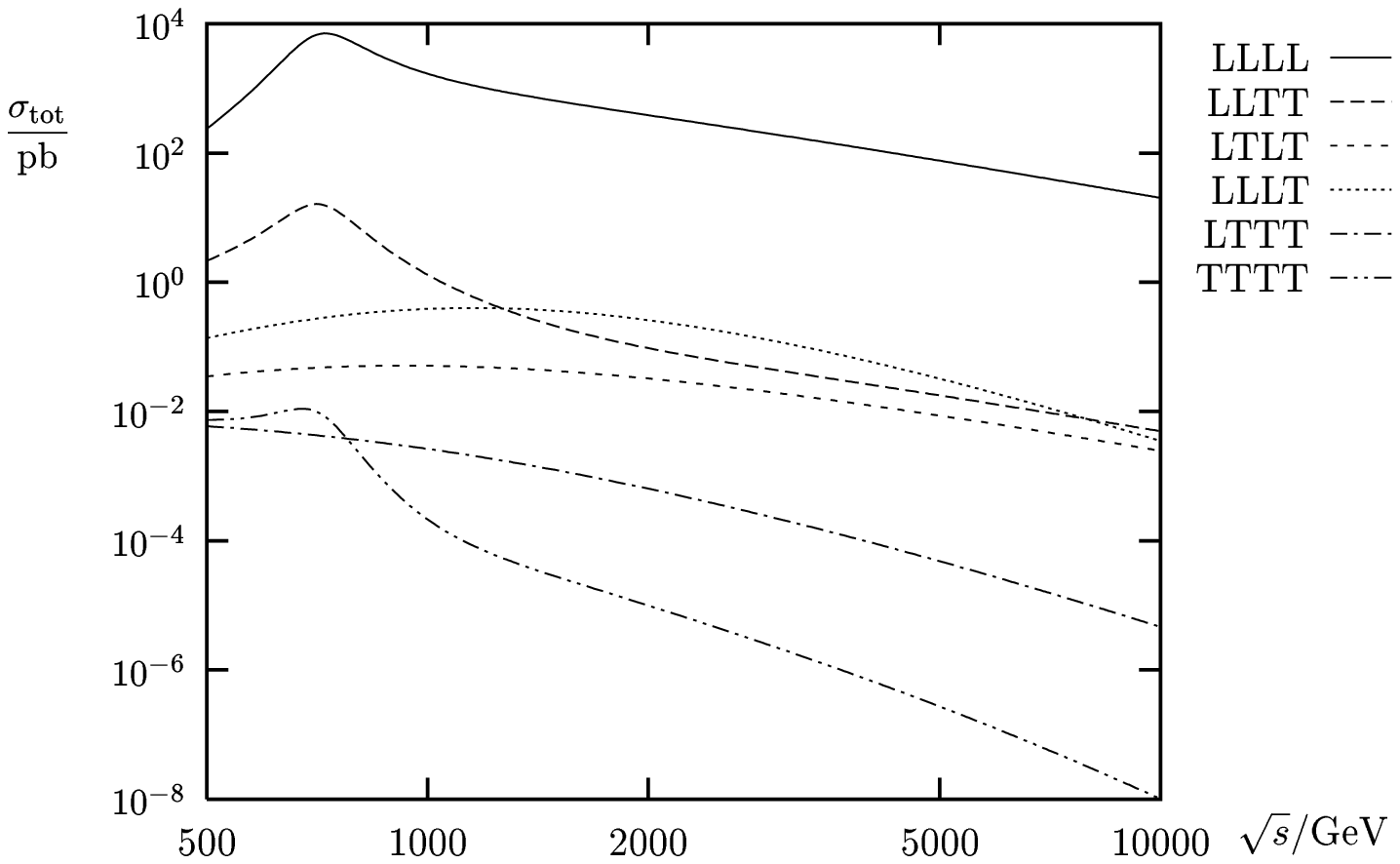}
\caption{Integrated lowest-order \css\ for various polarizations 
at $\MH=700\GeV$}
\label{fi:born.MH=700}
\end{figure}
The \cs\ for purely longitudinal
\PZ~bosons (LLLL) and the ones with two transverse and two longitudinal
\PZ~bosons (LLTT, LTLT) behave as $1/s$ for high energies.
The \css\ with one or three longitudinal gauge bosons (LTTT, LLLT) are
proportional to $1/s^2$, and the \cs\ for purely transverse
\PZ~bosons (TTTT) vanishes as $1/s^3$ at high energies
for $\theta_\cut>0$.
When integrated over the full scattering angle
the TTTT \cs\ behaves as $1/s^2$ owing to the $t$- and $u$-channel poles.
The \css\ not shown in \refta{ta:highenergy} 
are obtained from the symmetry relations
\refeq{boserel} and \refeqf{CPTrel}, \eg $\si_{\mathrm{LLLT}} = 
\si_{\mathrm{LLTL}} = 2\si_{\mathrm{TLLL}}$, where the factor 2 originates 
from the different spin average.

Figure~\ref{fi:born.MH=100} illustrates the 
(exactly calculated) polarized lowest-order \css\ integrated
over $\theta_\cut<\theta<180^\circ-\theta_\cut$ for
$\theta_\cut=10^\circ$ and a low Higgs-boson mass of $\MH=100\GeV$.
In \reffi{fi:born.MH=700} we show the same \css\ for $\MH=700\GeV$
using a \naive\ constant finite width to describe the 
Higgs-boson resonance,
as discussed in \refse{se:Hres} below.
The enhancement of the 
LLLL and LLLT \css\ caused by 
the factor $\MH^2/\MZ^2$ can be seen by comparing 
\reffis{fi:born.MH=100} and \ref{fi:born.MH=700}. 
The Higgs-boson resonance occurs only for equally polarized Z
bosons in the initial and final state, $\la_1=\la_2$, $\la_3=\la_4$.
At lowest order,
the LLLL \cs\ dominates independently of the Higgs-boson mass.

\section{Radiative corrections}
\label{se:rcs}

\subsection{Calculational framework}
\label{se:calframe}

We have performed the calculation of the radiative corrections 
(RCs)
in 't Hooft--Feynman gauge both in the conventional formalism 
and in the background-field formalism, applying the on-shell
renormalization scheme in both cases. We follow the conventions of 
\citere{adhab} for the conventional formalism 
and of \citere{bgflong} for the background-field formalism. 
In the conventional formalism the field renormalization 
is fixed such that no external wave-function
renormalization is needed. 
In the renormalization scheme introduced in \citere{bgflong} for
the background-field method the field
renormalization is determined by gauge invariance, and
a non-trivial external wave-function renormalization is required, as 
explicitly described in \citere{bgfet}.

The Feynman graphs have been generated and drawn 
with {\it FeynArts} \cite{FA}.
Both in the conventional formalism and in the background-field method we
have performed two independent calculations.

One evaluation is based on the calculational method described in 
\citere{adhab}. 
With the help of {\it Mathematica} \cite{math} the 
amplitudes 
are decomposed into SMEs and invariant functions, and the 
one-loop contributions to the
invariant functions are expressed in terms of standard tensor integrals. 
The tensor integrals
are reduced to the standard scalar one-loop integrals,
as described in \citere{Pa79}. 
The scalar
one-loop integrals are evaluated using the methods and general results of
\citere{tH79}. The last two steps are performed numerically using own {\it
Fortran} routines. 

In the other calculation the algebra 
is performed with {\it Mathematica}
and {\it Form} \cite{Form}
and has been partially checked with {\it FeynCalc} \cite{FC}. 
The resulting symbolic amplitudes are 
automatically converted into a {\it Fortran} 
program. Instead of using SMEs, 
all scalar products of four-vectors 
are grouped together and calculated at
run-time by inserting the explicit representations (\ref{polvecs}) for the
polarization vectors. 
The tensor integrals are numerically reduced to scalar
integrals, which are evaluated using the {\it FF\/} package \cite{FF}. 
The code thus obtained executes favorably fast and numerically stable.

Because of the length of the results we do not list the analytical
expressions but give only an inventory 
of the $\O(\alpha)$ RCs
and discuss some important features. 

\subsection{\boldmath{Inventory of $\O(\alpha)$ corrections}}

Both in the conventional formalism and in the background-field formalism
about 550 Feynman diagrams contribute to $\zzzz$ at one-loop order. 
The one-loop corrections can be classified into
self-energy corrections, vertex corrections, box corrections,
and wave-function-renormalization corrections.
All of them can be divided into $s$-,
$t$-, and $u$-channel contributions, 
which are related by simple transformations.
In the following we
list only the $s$-channel
Feynman graphs for the conventional
formalism in 't~Hooft--Feynman gauge.

\begin{figure}
\begin{center}
\epsfig{figure=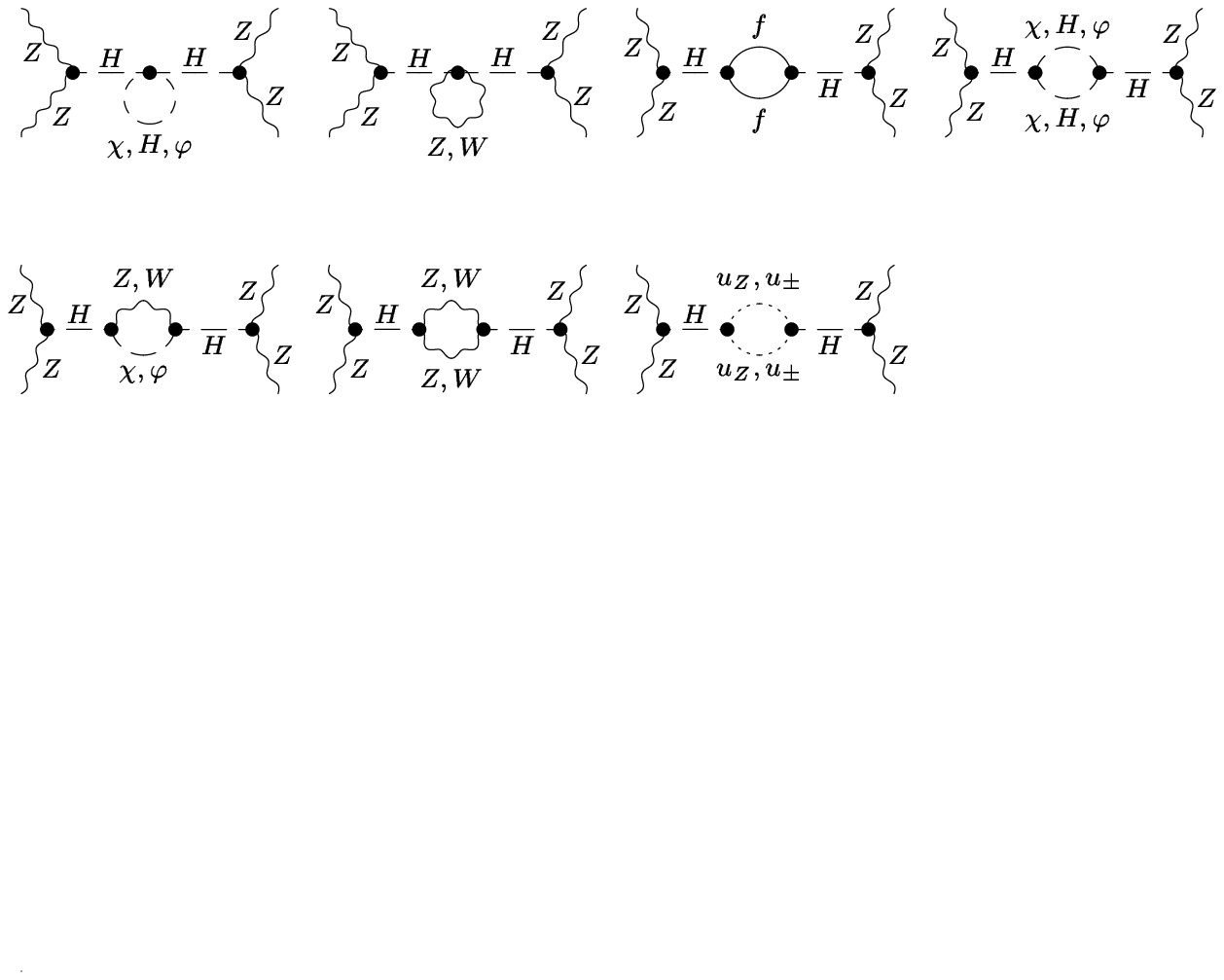}
\end{center}
\caption{\label{selfediags}$s$-channel self-energy diagrams}
\end{figure}

The diagrams contributing to the self-energy corrections in the $s$ channel 
are shown in \reffi{selfediags}. 
The diagram with a virtual $W$ and $\phi$ field 
actually represents two
diagrams with opposite charge flow.
After renormalization the diagrams of \reffi{selfediags}
yield together with the corresponding $t$- and $u$-channel diagrams
the following contribution to the invariant matrix element:
\begin{equation} \label{seamp}
\de\M _{\self}=\frac{e^2\MZ^2}{\cw^2\sw^2}\left(
\frac{\M _{00}^{(s)}}{(s-\MH^2)^2}\Sigma^{H}(s)+
\frac{\M _{00}^{(t)}}{(t-\MH^2)^2}\Sigma^{H}(t)+
\frac{\M _{00}^{(u)}}{(u-\MH^2)^2}\Sigma^{H}(u)\right) \,,
\end{equation}
where $\Sigma^{H}$ is the renormalized Higgs-boson self-energy.

\begin{figure}
\begin{center}
\epsfig{figure=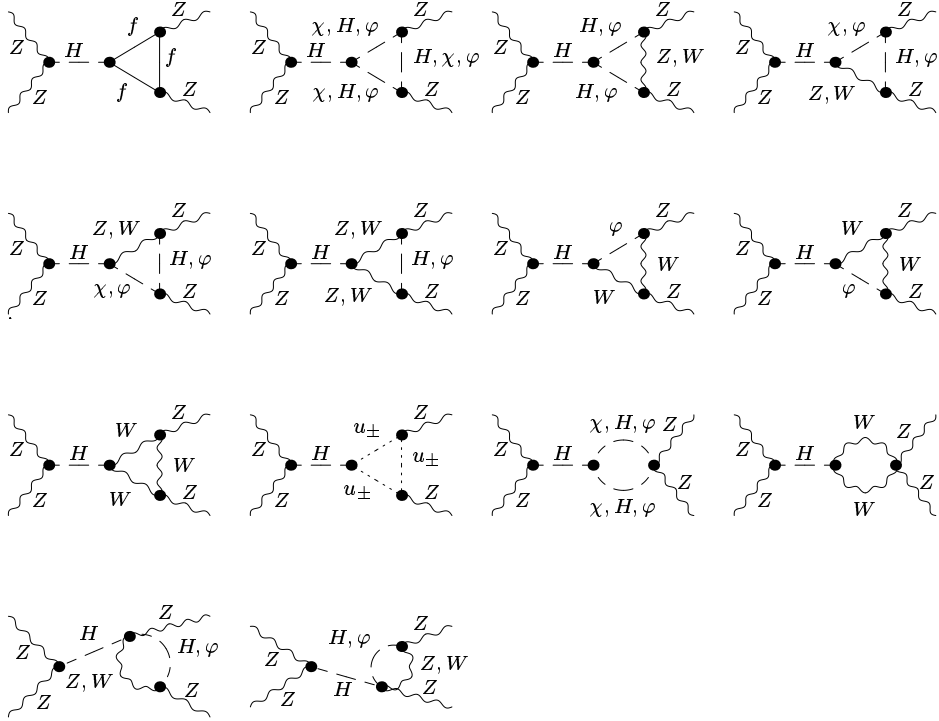}
\end{center}
\caption{\label{tridiags}$s$-channel final-state vertex diagrams}
\end{figure}

For each of the six vertices appearing in the Born diagrams (\reffi{borndiags})
there is a set of vertex corrections. In \reffi{tridiags} we show the 
diagrams that constitute the corrections to the 
final-state vertex in the $s$-channel diagram
of \reffi{borndiags}. 
Note that each graph with three charged fields
or two different charged fields
in the loop 
represents two diagrams with opposite orientation of the charge flow.
Owing to the simple tensor structure of the $ZZH$ vertex, the
vertex corrections have the relatively simple form
\beq \label{vertamp}
\de\M_{\vertex} = -\frac{e^2\MZ^2}{\cw^2\sw^2} \sum_{r=s,t,u}
\frac{1}{r-\MH^2} \left[ 2\M_{00}^{(r)}F_0^{ZZH}(r)
+ (\M_{10}^{(r)} + \M_{01}^{(r)}) F_1^{ZZH}(r) \right]
\eeq
with the two renormalized
form factors $F_0^{ZZH}(r)$ and $F_1^{ZZH}(r)$ for each channel,
the corresponding SMEs $\M_{00}^{(r)}$ from \refeq{eq:M00}, and
\beqar\label{vertexSME}
\M_{01}^{(s)} &=& 
(\veps_1\cdot\veps_2)(\veps_3^*\cdot k_4) (\veps^*_4\cdot k_3)/s , \qquad
\M_{10}^{(s)} = 
(\veps^*_3\cdot\veps^*_4)(\veps_1\cdot k_2) (\veps_2\cdot k_1)/s , \nl
\M_{01}^{(t)} &=& 
(\veps_1\cdot\veps^*_3)(\veps_2\cdot k_4) (\veps^*_4\cdot k_2)/s , \qquad
\M_{10}^{(t)} = 
(\veps_2\cdot\veps^*_4)(\veps_1\cdot k_3) (\veps^*_3\cdot k_1)/s , \nl
\M_{01}^{(u)} &=& 
(\veps_1\cdot\veps^*_4)(\veps_2\cdot k_3) (\veps^*_3\cdot k_2)/s , \qquad
\M_{10}^{(u)} = 
(\veps_2\cdot\veps^*_3)(\veps_1\cdot k_4) (\veps^*_4\cdot k_1)/s .
\label{eq:M01}
\eeqar
The factors $1/s$ in \refeq{eq:M01} 
have been introduced to render the matrix elements dimensionless.

\begin{figure}
\begin{center}
\epsfig{figure=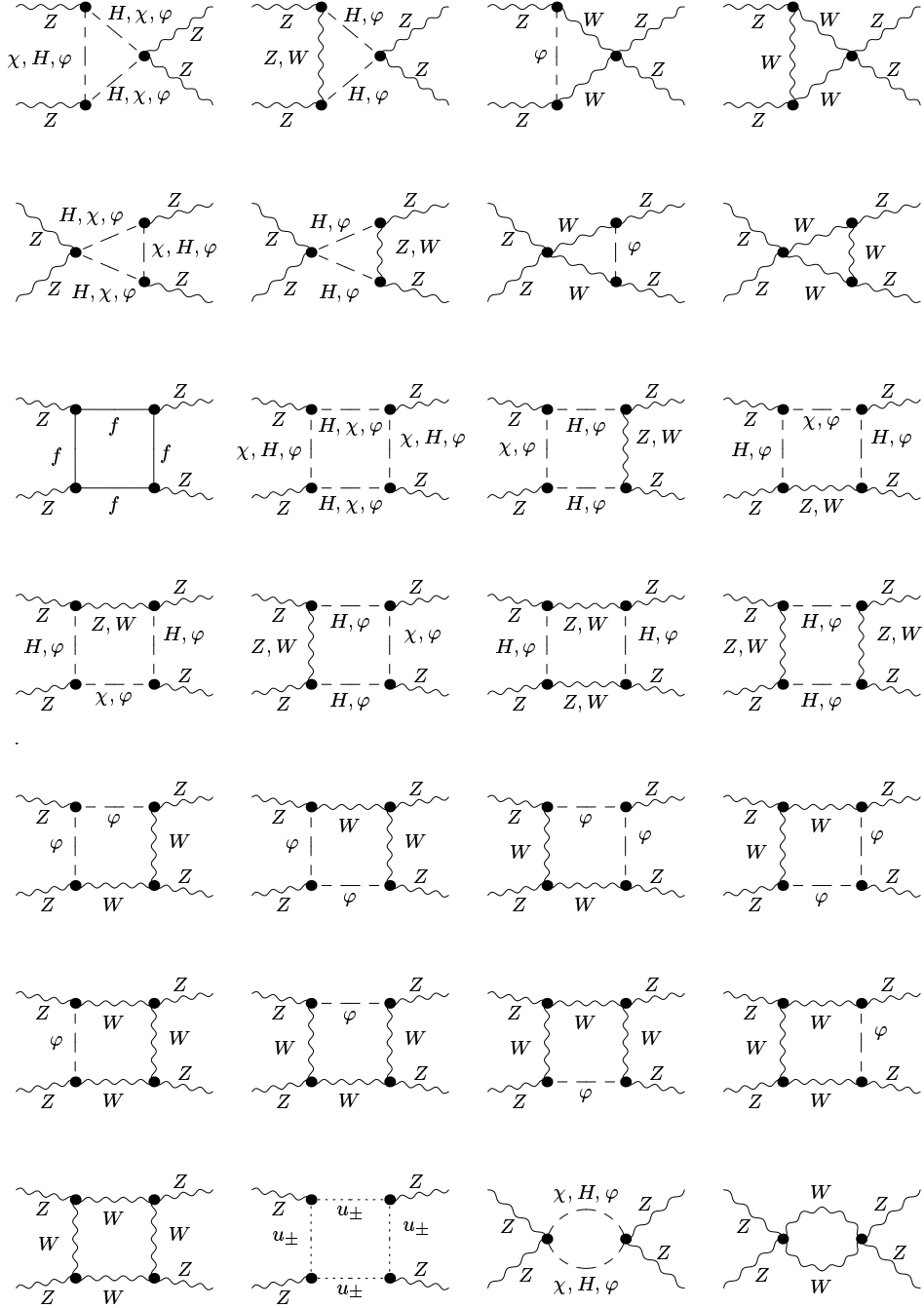}
\end{center}
\caption{\label{boxdiags}$s$-channel box diagrams}
\end{figure}

The $s$-channel box diagrams
(\ie those with natural variables $s$ and $t$)
are shown in \reffi{boxdiags}.
Note that again all 
graphs with three or four charged fields 
in the loop represent two Feynman diagrams with 
opposite orientation of the charge flow.
The analytical expressions for the box diagrams are rather involved and
require all CP-conserving SMEs. 
It turns out that the results for the bosonic box diagrams are 
shorter 
by a factor of about 3/2
in the background-field formalism
as compared to the conventional formalism.
Because of the involved structure of the fermion--\PZ-boson couplings,
the most complicated expressions are those for the fermionic box diagrams,
which are identical in both formalisms.

Following the complete on-shell renormalization scheme of \citere{adhab} 
for the conventional formalism the field renormalization is chosen such
that no extra wave-function renormalization is necessary. However, in
the on-shell renormalization scheme of \citere{bgflong} for the
background-field formalism a non-vanishing wave-function renormalization
for external particles
is required.
In a strict
$\O(\al)$ calculation the wave-function renormalization corrections are
given by
\beq
\de\M_{\wf} = 2\,\de R_Z\,\M_\Born,
\eeq
where
\beq
\de R_Z = - \Re \left\{\Si^{\prime ZZ}_{\mathrm{T}}(\MZ^2)\right\}
\eeq
is the $\O(\al)$ contribution to the wave-function renormalization constant 
$R_Z=1+\de R_Z$. The function 
$\Si^{\prime ZZ}_{\mathrm{T}}(s) = \rd \Si^{ZZ}_{\mathrm{T}}(s)/\rd s$
denotes the derivative of the transverse part of the
renormalized Z-boson self-energy.

Thus, the full one-loop matrix element reads
\beq
\de\M_{\oneloop} = \de\M_\self + \de\M_{\vertex} 
+ \de\M_{\boxrc} + 
\de\M_{\wf}.
\eeq

\subsection{Corrected \cs}

It turns out that the $\Oa$ corrections are comparable or even larger
than the lowest-order contributions for various important configurations.
In order to obtain meaningful predictions,
it is therefore necessary to consider not only the interference
between the lowest-order and the one-loop matrix element but 
to take into account the complete square of the matrix element and 
to define the corrected \cs\ as 
\beq\label{sigmacorr}
\frac{\rd\sigma }{\rd\Omega}=\frac 1{64\pi^2 s}
\left| \M_\born + \de\M_{\oneloop} \right|^2
\eeq
including the terms involving $\left|\de\M_{\oneloop}\right|^2$.

In this way we end up with $\O(\al)$ accuracy 
where $\M_\born$ dominates and $\O(1)$ accuracy otherwise
(relative to the leading loop order).
To obtain $\O(\al)$ accuracy everywhere
the $\O(\al^2)$ corrections would be required. Note that the
interference between $\de\M_{\twoloop}$ and $\M_\Born$ is suppressed
with respect to $|\de\M_{\oneloop}|^2$ 
if the importance of $\de\M_{\oneloop}$ results from a suppression of
$\M_\Born$ and not from the presence of an effective  large loop-expansion
parameter.

\begin{figure}
\epsfig{figure=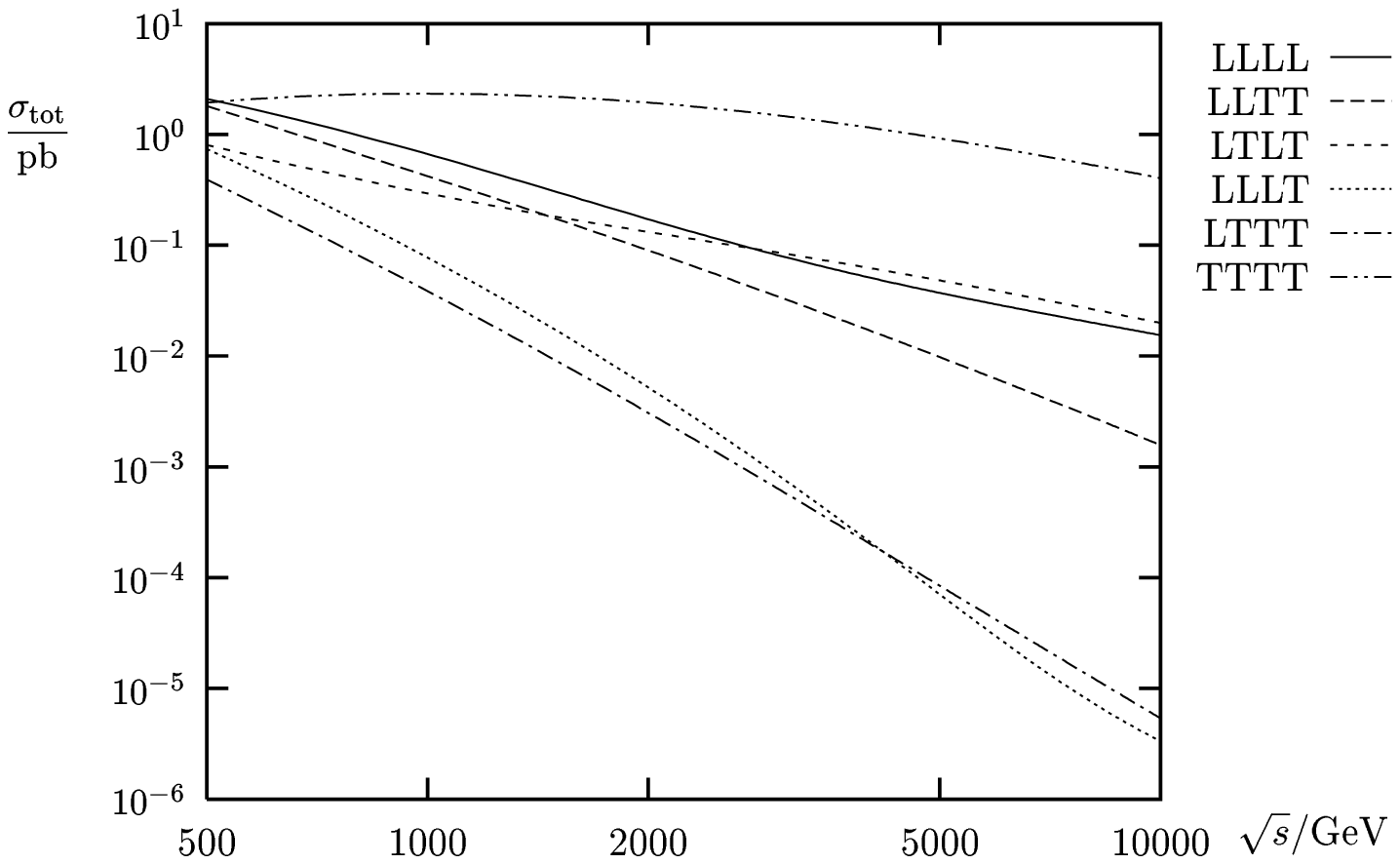}
\caption{Integrated corrected \css\ for various polarizations
at $\MH=100\GeV$}
\label{fi:onel.MH=100}
\vspace{2em}
\epsfig{figure=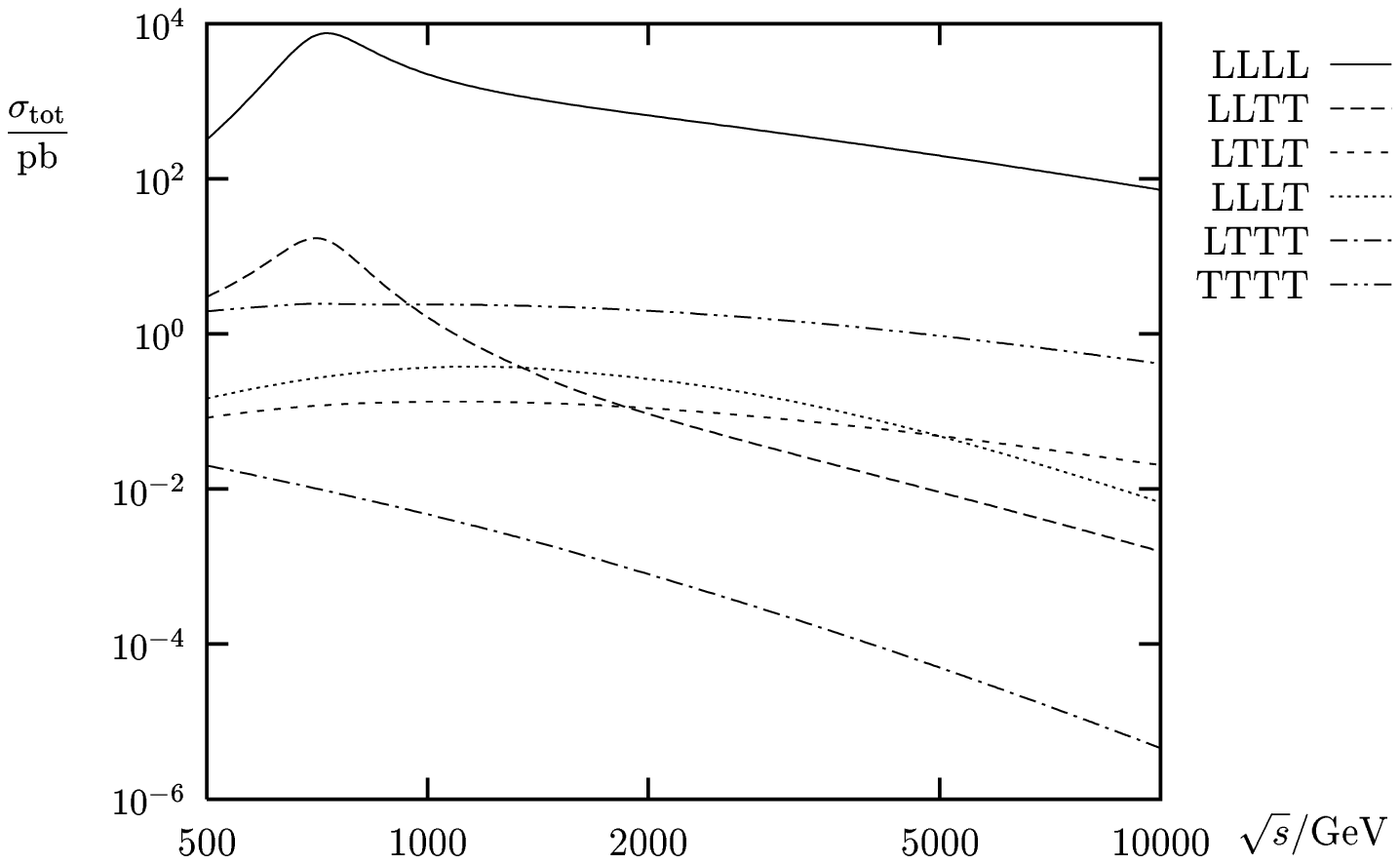}
\caption{Integrated corrected \css\ for various polarizations
at $\MH=700\GeV$}
\label{fi:onel.MH=700}
\end{figure}

In \reffis{fi:onel.MH=100} and \ref{fi:onel.MH=700} we show the
corrected polarized \css\ as defined in \refeq{sigmacorr}.
The Higgs-boson resonance 
in \reffi{fi:onel.MH=700}
is treated by Dyson summation within the BFM [\cf
\refse{se:BFMdyson}, \refeq{MDyson}].
In the case of purely transverse 
\PZ~bosons 
(TTTT) the \cs\ is drastically enhanced compared to the
lowest order and behaves as $1/s$ at high energies (without cut-off it
even would not go down with energy). 
For a small Higgs-boson mass this \cs\ becomes the dominating one.
Apart from the LTTT case, the corrections reach the size of the
lowest-order \css\ for all polarizations.
This is probably due to corrections of the form%
\footnote{At high energies vertex and box corrections typically yield
contributions of this kind, as e.g.\ explicitly calculated for
$\Pep\Pem\to\PWp\PWm$ in \citere{be93}.}
$(\alpha/\pi)(\ln(s/\MZ^2))^2 \approx 0.2$ that are further enhanced by
numerical factors and $t$- and $u$-channel poles.
The relative corrections depend only weakly on the Higgs-boson mass
apart from the polarizations LLLL and LLLT, where the corrections
involve extra factors $\MH/\MZ$.
As a consequence the LLLL \cs\ dominates for a large Higgs-boson mass.
As for the lowest-order cross-sections,
the Higgs-boson resonance only
contributes for $\la_1=\la_2$ and $\la_3=\la_4$. For the TTTT channel
the resonance is proportional to the corresponding strongly suppressed
Born cross-section and thus not visible. 

Because the lowest-order \css\ are not dominating, the
universal corrections associated with the running of $\al$ and the $\rho$
parameter, which are related to the lowest order,  are not leading.

\subsection{Landau singularities in four-point functions}
\label{subse:sing}

The four-point function \cite{tH79} 
\begin{equation}
D_0(\MZ^2,\MZ^2,\MZ^2,\MZ^2,t,u,m^2,m^2,m^2,m^2)\,,
\label{eq:d0}
\end{equation}
exhibits a Landau singularity \cite{d0pol} of the form
\begin{equation}
\label{landausing}
D_0=D_0^{\reg}-\frac{\pi ^2}{\sqrt{\mathstrut
\Delta -\ri\epsilon }}
\end{equation}  
for $t<0$, $u<0$, and $\MZ>2m$,
where $D_0^{\reg}$ is regular and 
\begin{equation}
\Delta =\frac{tu}{16}\left[ \left(t-4m^2\right) \left(u-4m^2\right) 
-\left(2\MZ^2-4m^2\right)^2\right] \,.
\end{equation}
With $b=\MZ^4-m^2s$ and $p^2=E^2-\MZ^2$ this becomes
\begin{equation}
\Delta =(p^4\sin ^2\theta -b)p^4\sin ^2\theta\,.
\end{equation}

Squaring the matrix element
promotes the root singularity at $p^2\sin\theta = \pm \sqrt{b}$ 
to a pole which is not integrable and thus leads to a formally divergent \cs. 

This singularity should disappear from
physical observables. The condition $\MZ>2m$ suggests that 
it is related to the instability of the \PZ~bosons.
In fact, as illustrated in \refapp{app:d0sing}, it is canceled 
by diagrams that contribute to the inclusive process $\PZ\PZ\to 4\Pf$, 
which cannot be separated from $\zzzz$ once the
decay of the \PZ~bosons is taken into account.
Moreover, one should notice that colliding Z bosons which are radiated
off from incoming particles possess an invariant mass $q^2<0$ so that
the condition $q^2>4m^2$ is never 
fulfilled in the physical region of
phase space. The use of on-shell Z bosons ($q^2=\MZ^2$) is just part of
the equivalent vector-boson approximation.

\begin{figure}
\begin{center}
\epsfig{figure=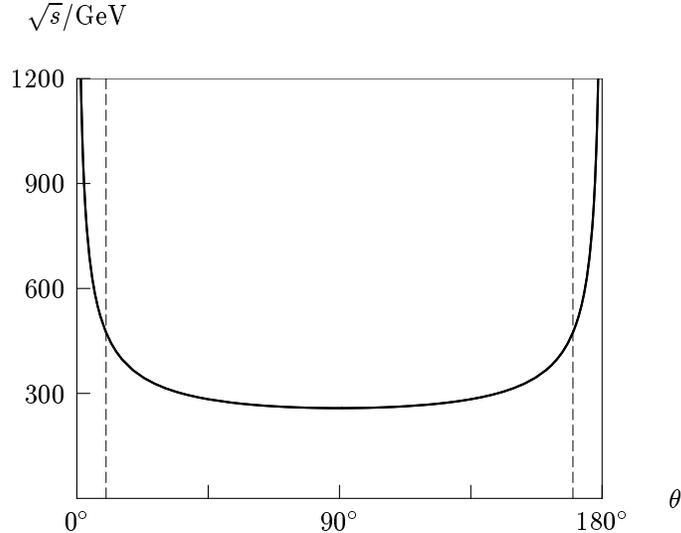}
\end{center}
\caption{\label{d0plot}The location of the Landau singularities in the $D_0$
function \refeq{eq:d0} for $m=0$. 
The dashed lines at $\theta _{\hbox{\scriptsize cut}}=10^\circ$ and $180^\circ
-\theta _{\hbox{\scriptsize cut}}=170^\circ$ indicate the integration interval.}
\end{figure}

The Landau
singularity appears in practice for box diagrams involving light fermions,
\ie with $m=\Mf\ll\MZ$.  
The location of the singularity in phase space for $m=0$ 
is shown in \reffi{d0plot}.
It appears at $\theta= 90^\circ$ for $p^2=\MZ^2$, \ie
$\sqrt{s}=2\sqrt{2}\MZ\approx258\GeV$,
and moves fast towards the 
forward and backward directions with increasing energy.
Its effect is most prominent at low energies and becomes small at high
energies. Moreover, it is located outside the angular
region $10^\circ<\theta<170^\circ$ 
for $\sqrt{s}\gsim500\GeV$ for all fermions.
In the following we always consider the \cs\ 
in regions where this singularity is absent or negligible.

\section{Higgs-boson resonance}
\label{se:Hres}

Diagrams that involve a 
Higgs-boson propagator in the $s$ channel have a pole
at $s=\MH^2$.
The double pole in the self-energy diagrams of \reffi{selfediags} is
reduced to a single pole after Dyson summation.

If $\MH>2\MZ$ a proper
treatment of the resonance is necessary to render the 
\cs\ finite and meaningful. 
The \naive\ introduction of a finite Higgs-boson
width via the substitution 
\begin{equation}
\frac 1{s-\MH^2}\longrightarrow \frac 1{s-\MH^2+\ri \MH \Ga_\PH}
\end{equation}
or \naive\ Dyson summation 
\begin{equation}
\frac 1{s-\MH^2}\longrightarrow \frac 1{s-\MH^2+\Sigma^H(s)}
\end{equation}
amounts to an inclusion of an incomplete set of higher-order
corrections such that the resulting matrix element becomes 
gauge-dependent and violates the Ward identities and thus 
also the gauge cancellations, which guarantee unitarity.

\subsection{Pole expansion}
\label{se:poleexpansion}

Since the poles of the $S$ matrix are gauge-independent,
it has been proposed \cite{St91} to perform a Laurent expansion
about the complex pole. In a \naive\ way 
this can be done by decomposing the contributions of the resonant
diagrams into resonant and non-resonant parts and introducing the finite
width only in the former.
For vertex corrections this leads to the substitution
\def\sp{s_p}
\def\coup{\frac{e^2\MZ^2}{\cw^2\sw^2}}
\beqar \label{resonancetreatment1}
\de\M_{\vertex}^{(s)} &=& -\coup
\sum_{i=0,1} (\M_{i0}^{(s)} + \M_{0i}^{(s)})\frac{F^{ZZH}_i(s)}{s-\MH^2} \nl
&\to& -\coup
\sum_{i=0,1} (\M_{i0}^{(s)} + \M_{0i}^{(s)})
\left[\frac{F^{ZZH}_i(\MH^2)}{s-\MH^2+\ri\MH \Ga_\PH}
+\frac{F^{ZZH}_i(s)-F^{ZZH}_i(\MH^2)}{s-\MH^2} \right] ,
\nn\\
\eeqar
where [to $\Oa$ accuracy]
\beq
\Gamma_H=\Im \Sigma^H(\MH^2)/\MH
\label{eq:GH}
\eeq
is the decay width of the Higgs boson.
For the lowest-order and self-energy contributions we write
\beqar \label{resonancetreatment2}
\M_\Born^{(s)} + \de\M_{\self}^{(s)} &=& -\coup
\M_{00}^{(s)}\frac{1}{s-\MH^2}\left[1-\frac{\Si^H(s)}{s-\MH^2}\right] 
\nl
&=& -\coup
\M_{00}^{(s)}\frac{1}{s-\MH^2}\biggl[
1-\frac{\Si^H(\MH^2)}{s-\MH^2} - \Si^{\prime H}(\MH^2) \nl
&& \qquad\qquad{}
-\frac{\Si^H(s)-\Si^H(\MH^2)-(s-\MH^2)\Si^{\prime H}(\MH^2)}{s-\MH^2}
\biggr]\nl
&\to& -\coup
\M_{00}^{(s)}\biggl[\frac{1}{s-\MH^2+\ri\MH \Ga_\PH}
\left(1-\Si^{\prime H}(\MH^2) \right) \nl
&& \qquad\qquad{}
-\frac{\Si^H(s)-\Si^H(\MH^2)-(s-\MH^2)\Si^{\prime H}(\MH^2)}{(s-\MH^2)^2}
\biggr]
\eeqar
with $\Si^{\prime H}(s) = \rd \Si^{H}(s)/\rd s$. We have used 
the fact that the renormalized 
Higgs-boson self-energy fulfills $\Re\Si^{H}(\MH^2) = 0$ and that 
$\Im \Sigma^H(\MH^2)=\MH\Gamma _H $ is accounted for by the resummed
terms.
As the inclusion of a finite width corresponds just to the inclusion of
higher-order terms, all expressions 
are equivalent at the one-loop level.

According to \refeq{eq:GH}, $\Gamma_H$ is a tree-level quantity if
$\Sigma^H$ is calculated at one-loop order. Consequently, in a one-loop
calculation one ends up with ${\cal O}(1)$ accuracy on resonance
and different treatments of the finite \PZ-boson width differ relatively
in $\Oa$. 
An improvement to $\Oa$ accuracy on
resonance would require to use the fully $\Oa$-corrected Higgs-boson
width in the resonant denominators.

The above procedure is gauge-independent because we modify the amplitude
by terms that depend only on the gauge-independent residue of the pole and 
the physical mass and width.
However, the actual application of the pole expansion deserves some care.
Firstly, the above treatment is not uniquely determined by the resonance pole,
because the Laurent expansion is only applied to the form factors but not 
to the SMEs and the split-up between these two is not unique.
On the other hand, the terms introduced by the
modification of the amplitude in general 
violate the Ward identities and thus eventually unitarity.
This problem could be avoided and the pole scheme could be uniquely defined 
by including the complete matrix elements into the
Laurent expansion. This, however, leads to problems in defining 
the residues, \ie in particular the corresponding momenta and wave functions,
for more general processes
in certain kinematical regions (\cf \citere{Ae93}).
In the following we show how one can exploit the above-mentioned
freedom in the pole expansion in order to 
eliminate unitarity-violating terms.

We first illustrate the procedure at tree level.
A general pole expansion is obtained by absorbing some 
arbitrary function $f(s)$ with $f(\MH^2)=1$ into the 
SME $ \M_{00}^{(s)}$ and performing the
Laurent expansion for the resulting modified form factor.
After resubstituting the original SME this amounts to the replacement
\beqar 
\label{modborn}
&&\M_{\Born}^{(s)} 
\to -\coup \M_{00}^{(s)}
\left[\frac{f(s)}{s-\MH^2+\ri\MH \Ga_\PH} +\frac{1-f(s)}{s-\MH^2} \right].
\eeqar
The added terms are proportional to
$f(s)\M_{00}^{(s)}\Ga_\PH/(s-\MH^2)/(s-\MH^2+\ri\MH\Ga_\PH)$.
If the matrix element 
$f(s)\M_{00}^{(s)}$ grows too fast with energy these 
terms violate unitarity at high energies.
In the high-energy limit the ratio between the Born cross-sections 
for longitudinal Z bosons (LLLL)
calculated for $\Ga_\PH\ne0$ and $\Ga_\PH=0$ behaves as 
$1+f(s)^2\Ga_\PH^2/9\MH^2$ for real $f(s)$, 
\ie $f(s)=1$ yields a result that is off
by a constant factor, but for instance $f(s)=\MH^2/s$ reproduces the
correct high-energy limit.
While different choices of $f(s)$ by construction do not modify the
resonant contribution they differ evidently in the non-resonant terms.
This indefiniteness of the non-resonant lowest-order contributions
gives rise to ambiguities of relative $\Oa$ in the resonance region.

At the one-loop level the generalized pole expansion is 
obtained by absorbing arbitrary functions $f_{ij}(s)$ with
$f_{ij}(\MH^2)=1$ into the SMEs $\M_{ij}^{(s)}$ before performing the
Laurent expansion of the form factors. Besides the appearance of several
functions $f_{ij}$ such a general pole expansion even includes terms
involving their derivatives $f'_{ij}$. For our purposes it is sufficient
to consider the following modified pole expansion
\beqar 
\de\M_{\vertex}^{(s)} 
&\to& 
-\coup \sum_{i=0,1} (\M_{i0}^{(s)} + \M_{0i}^{(s)})
\nl*
\label{resonancetreatment3}
&& \hspace{1.0cm}{} \times
\left[\frac{F^{ZZH}_i(\MH^2)f(s)}{s-\MH^2+\ri\MH \Ga_\PH}
+\frac{F^{ZZH}_i(s)-F^{ZZH}_i(\MH^2)f(s)}{s-\MH^2} \right],\  \nl
\M_\Born^{(s)} + \de\M_{\self}^{(s)} 
&\to& 
-\coup \M_{00}^{(s)}\biggl[\frac{1}{s-\MH^2+\ri\MH \Ga_\PH}
\left(1-\Si^{\prime H}(\MH^2)f(s) \right)\nl
&&\hspace{2.0cm}{}
- \frac{\Si^H(s) -\Si^H(\MH^2)
-(s-\MH^2)\Si^{\prime H}(\MH^2)f(s)}{(s-\MH^2)^2}
\biggr],
\eeqar
which differs from a consistent expansion with  $f_{ij}(s)=f(s)$ only
by terms of the order of $\Gamma_\PH^2$, \ie $\Oaa$. 
If we  introduce the finite width as in \refeq{resonancetreatment1}
and \refeqf{resonancetreatment2}, \ie with $f(s)=1$,
we modify the \cs\ for longitudinal 
gauge bosons at high energies by a constant contribution
of $\O(\al\Ga_\PH/\MH)$.
In the modified version of the pole
scheme the high-energy behavior can be improved by choosing a suitable
function $f(s)$ that vanishes sufficiently fast at high energies. With
our definition of the SMEs \refeq{vertexSME} it is sufficient to choose 
$f(s) = \MH^2/s$.
Note that if we did not include factors $\MH^2/s$ to render the SMEs 
dimensionless
we would obtain a contribution to the matrix element that grows with $s$
for $f(s)=1$, \ie that violates unitarity.
The freedom parametrized by $f(s)$ in \refeq{resonancetreatment3}
affects the non-resonant contributions in $\Oa$. 
On resonance this introduces ambiguities in $\Oaa$
relative to the leading resonant terms.

The above recipe for 
the usual on-shell renormalization scheme 
is directly connected 
with an expansion of the transition matrix element about its complex pole. 
In \citere{St91} such an expansion was explicitly described 
for angular-independent resonances, where the complications \cite{Ae93}
in defining wave functions and momenta on resonance are absent.
The procedure of \citere{St91} can be directly transfered to $\zzzz$.
In this respect only the angular-independent, 
\ie the 
(one-particle-reducible) lowest-order, self-energy, and
vertex contributions in the $s$-channel, $\M^{(s)}_{\onePR}$, are relevant.
After Dyson summation these can be written (assuming no truncation of the
perturbation series) as follows:
\beq\label{eq:Ms}
\M^{(s)}_{\onePR} =
-\coup \sum_{i,j=0,1} \M_{ij}^{(s)}
\frac{(\de_{0i}+ F^{ZZH}_i(s))(\de_{0j}+ F^{ZZH}_j(s))}
{s-\MH^2+\Si^H(s)},
\eeq
\ie as a product of full vertex functions and the full propagator.
The additional SME $\M_{11}^{(s)}$ is defined as
\beq
\M_{11}^{(s)} = (\veps_1\cdot k_2) (\veps_2\cdot k_1)
(\veps_3^*\cdot k_4) (\veps^*_4\cdot k_3)/s^2.
\eeq
The complex pole $\sp$ of \refeq{eq:Ms} is obtained as the solution of
\beq
\sp - \MH^2 +\Sigma^H(\sp) = 0.
\eeq
Since $\M^{(s)}_{\onePR}$ is 
analytical in $s$, it can be continued to complex $s$
and expanded about $s_p$. 
The leading term in this expansion is given by the 
resonant part 
\beq\label{Monshell}
\M_{\mathrm{reso}} =
-\coup \sum_{i,j=0,1} \left.\M_{ij}^{(s)}\right|_{s=\sp} 
\frac{(\de_{0i}+ F^{ZZH}_i(s_p))(\de_{0j}+ F^{ZZH}_j(s_p))}
{(s-s_p)(1+\Si^{\prime H}(\sp))}. 
\eeq
The residue of $\M_{\mathrm{reso}}$ at the pole $1/(s-\sp)$
can be interpreted 
as the product of two physical amplitudes $\M^{H\to ZZ}$ for the
decay $\PH\to\PZ\PZ$.
To one-loop accuracy $\M^{(s)}_{\onePR}$ can be replaced by
\beqar\label{complexpoleexpansion}
\M^{(s)}_{\onePR} 
&\to&  -\coup
\sum_{i=0,1} (\M_{i0}^{(s)} + \M_{0i}^{(s)})
\left[\frac{F^{ZZH}_i(\sp)}{s-\sp}
+\frac{F^{ZZH}_i(s)-F^{ZZH}_i(\sp)}{s-\sp} \right] \\ 
&& -\coup
\M_{00}^{(s)}\frac{1}{s-\sp}\biggl[ 1 - \Si^{\prime H}(\sp)  
- \frac{\Si^H(s) -\Si^H(\sp)-(s-\sp)\Si^{\prime H}(\sp)}{s-\sp}
\biggr]. \nn
\eeqar
Owing to $\sp=\MH^2-\ri\MH\Ga_\PH+\Oaa$, the right-hand side of 
\refeq{complexpoleexpansion} 
differs from \refeq{resonancetreatment1} and
\refeqf{resonancetreatment2} only by higher-order contributions.

Note that the freedom in splitting the matrix element into SMEs and form
factors is also present in this approach, i.e.\ in
\refeq{complexpoleexpansion} we could also introduce functions 
$f_{ij}(s)$, as described in the first part of this section. 
In the considered case one could avoid this ambiguity by expanding
also the SMEs occuring in $\M^{(s)}$ about $s_p$. 
However, in more complicated situations it is not always possible to include
the wave functions in the pole expansion. If one then excludes the 
SMEs from the pole expansion, 
as for instance advocated in \citere{St96}, one 
is again confronted with the problem of violating Ward identities.

\subsection{Dyson summation within the background-field method}
\label{se:BFMdyson}

A different approach to introduce a finite width 
near resonances is to Dyson-sum the 
self-energy corrections.
It is a well-known fact that in the conventional formalism Ward
identities, which in particular rule the gauge cancellations, are
violated if Dyson summation is applied. However, in 
\citere{bgfet} it 
has been
shown that Dyson summation within the background-field method (BFM)
(see \citere{bgflong} and references therein) does not
violate the Ward identities if all 
one-particle-irreducible corrections
are taken into account in the same loop order.
Dyson summation naturally
arranges the reducible parts of 
amplitudes in a way that results from
forming trees with
vertex functions joined by full propagators (inverse two-point functions).
For the process under consideration this 
simply amounts to writing the
(one-particle-reducible)
lowest-order, self-energy, and vertex contributions in the following 
way:
\beqar\label{MDyson}
&& \M_\Born + \de\M_\self + \de\M_{\vertex}
\nl
&&\to \M_{\onePR} = 
-\coup \sum_{i,j=0,1} \sum_{r=s,t,u}\M_{ij}^{(r)} 
\frac{(\de_{0i}+ F^{ZZH}_i(r))(\de_{0j}+ F^{ZZH}_j(r))}{r-\MH^2+\Si^H(r)} , 
\eeqar
where $\Si^H$ and 
$F^{ZZH}_i$ denote the renormalized self-energy and form factors,
respectively,
in the BFM and
\beq
\M_{11}^{(t)} = (\veps_1\cdot k_3) (\veps_2\cdot k_4)
(\veps_3^*\cdot k_1) (\veps^*_4\cdot k_2)/s^2, \qquad
\M_{11}^{(u)} = (\veps_1\cdot k_4) (\veps_2\cdot k_3)
(\veps_3^*\cdot k_2) (\veps^*_4\cdot k_1)/s^2.
\eeq
The $s$-channel part of \refeq{MDyson} is formally identical 
to \refeq{eq:Ms}. Note, however, that we use \refeq{MDyson} in the
following with form factors and self-energy in finite, \ie one-loop, order
of perturbation theory.

The complete one-loop matrix element is obtained by adding the 
(one-par\-ticle-ir\-re\-du\-cible) box contributions and multiplying 
everything with the (UV-finite)
wave-function renormalization factor
$\sqrt{R_Z}$
for each external 
\PZ~boson,
\beq \label{MBFM}
\M_{\BFM} = (\M_{\onePR} + \M_{\boxrc}) \left(R_Z\right)^2.
\eeq
Since the wave-function renormalization factor $R_Z$ multiplies the 
complete matrix element, we can simply use its $\Oa$ approximation
\beq
(R_Z)^2 = 1-2\Re\Si_{\mathrm{T}}^{\prime ZZ}(\MZ^2).
\eeq

The matrix element \refeq{MBFM} fulfills all relevant Ward identities,
and, in particular, does not violate unitarity at high energies.
How\-ever, it depends on a gauge parameter via
higher-order (at least two-loop) corrections 
which are not completely taken into account.
This is nothing but a result of the fact that a Dyson summation
is always arbitrary to some extent.
Note also that the on-resonance self-energy is unique and equal to the
physical quantity $\ri\MH\Ga_\PH$.

In the conventional formalism the matrix element after Dyson summation
depends not only on the gauge but in addition on the choice of the
field renormalization. In the BFM the matrix element is actually
independent of the field renormalization. This can be seen as follows:
The field renormalization is fixed by background field gauge invariance
up to a UV-finite linear transformation of the renormalized fields.
Such a linear transformation turns the linear Ward identities for
the background-field vertex functions into Ward idenities for
transformed vertex functions with the same structure. These modified
Ward identities are still exactly valid even for full, \ie Dyson-summed,
propagators.
However, the effects of the linear transformation cancel
in $S$-matrix elements, thus giving a unique answer.'

The resonant part of the
Dyson-summed one-loop matrix element \refeq{MDyson} in the BFM reads
for $s=\MH^2$
\beq\label{MDysonres}
\left.\M_{\BFM}
\vphantom{\frac 12}
\right|_{s=\MH^2} \approx
-\coup (R_Z)^2 \sum_{i,j=0,1} \M_{ij}^{(s)}
\frac{(\de_{0i}+ F^{ZZH}_i(\MH^2))(\de_{0j}+ F^{ZZH}_j(\MH^2))}{\ri\MH\GH}.
\eeq
This differs from the resonant part of the pole-scheme amplitude
\refeq{resonancetreatment3},
\beq\label{Mpoleres}
\left.\M_{\pole}
\vphantom{\frac 12}
\right|_{s=\MH^2} \approx
-\coup \frac{\M_{00}^{(s)}(1-\Sigma^{\prime H}(\MH^2)) +
\sum_{i=0,1} (\M_{i0}^{(s)} + \M_{0i}^{(s)})
F^{ZZH}_i(\MH^2)}{\ri\MH\GH},
\eeq
in $\Oa$ in relative terms,
in accordance with the discussion after \refeq{eq:GH}.
Moreover, \refeq{MDysonres} is even gauge-dependent in 
this order on resonance, whereas \refeq{Mpoleres} is manifestly
gauge-independent.
The bulk of these effects can be attributed to the contribution of 
$\Si^{\prime H}(\MH^2)$ and thus to the different wave-function renormalization
in the BFM. 

In order to obtain the cross-section on resonance also in $\Oa$ accuracy
the imaginary part of the Higgs-boson self-energy has to be included in
two-loop order.
In the pole scheme this is equivalent to the introduction of the 
$\Oa$-corrected Higgs-boson width in the propagator. 
However, in the BFM approach all two-loop 
corrections are required in order to preserve the Ward identities.

\section{Cross-section for longitudinal Z bosons from the equivalence theorem}
\label{se:zzzzL}

The corrections of \OaMH\ to longitudinal
gauge-boson scattering processes have been
calculated in the literature \cite{Pa85,Da89,Gu93,VVVVcorr} 
using the equivalence theorem (ET) \cite{et,Ch85,Ya88,Gr95,bgfet}. 
The ET relates amplitudes involving longitudinal gauge
bosons with those involving the associated would-be Goldstone bosons
in the high-energy limit. Because the latter amplitudes are much easier
to be calculated, the ET was frequently used
to obtain \css\ in the high-energy limit.

\subsection{Equivalence theorem within the background-field method}

When using the ET in higher-order calculations one has to be careful to
include all correction factors that result from renormalization and
amputation \cite{Ya88}. 
It has been found in \citere{bgfet} that this is
particularly easy within the BFM.
In this formalism the
matrix elements for external longitudinal vector bosons are directly
obtained from the amputated Green functions with the corresponding
would-be Goldstone boson fields multiplied with the wave-function
renormalization constants of the gauge bosons.
Moreover, in contrast to the conventional formalism, in the BFM 
the ET is valid with and without Dyson summation.

\subsection{\boldmath{Results for $\zzzzL$}}

We want to apply the ET to the process $\PZ_\rL\PZ_\rL\to\PZ_\rL\PZ_\rL$
in the framework of the BFM
and investigate the accuracy of the corresponding predictions.

\begin{figure}
\begin{center}
\epsfig{figure=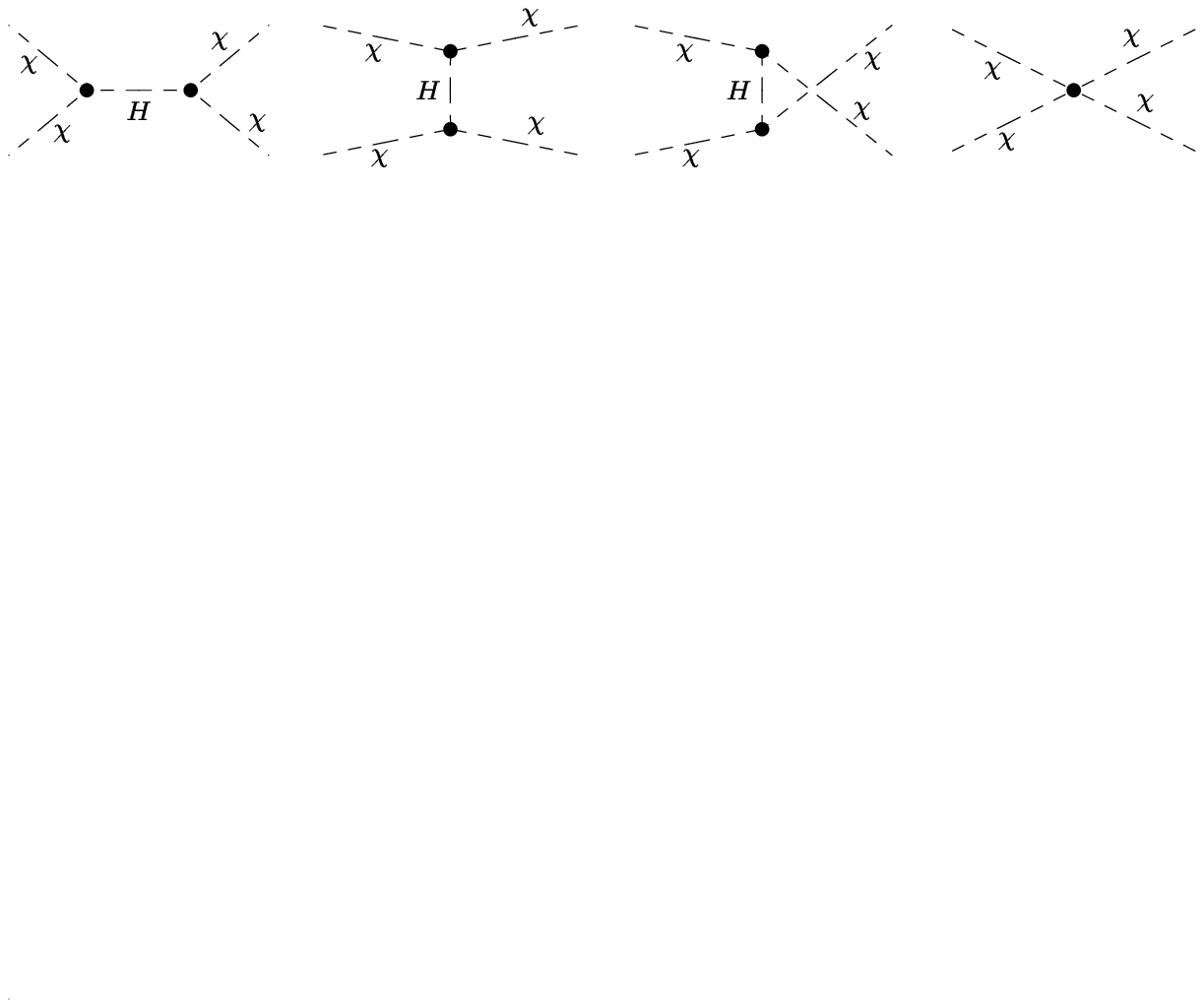}
\end{center}
\caption{\label{xxxxborndiags}Lowest-order diagrams for $\xxxx$ }
\end{figure}

To this end we have to consider the process $\chi\chi\to\chi\chi$, where
$\chi$ is the would-be Goldstone boson associated with the \PZ~boson.
In lowest order the four diagrams of \reffi{xxxxborndiags}
yield
\beq \label{xxxxbornamp}
\M^{\xxxx}_\born = -\frac{e^2\MH^2}{4\sw^2\cw^2\MZ^2}
\left[3 + \frac{\MH^2}{s-\MH^2} + \frac{\MH^2}{t-\MH^2} 
+ \frac{\MH^2}{u-\MH^2}\right].
\eeq
For $s\gg\MZ^2$ the matrix element for $\zzzzL$ approaches the one for
$\xxxx$ for any value of $\MH$ \cite{Gr95}.
We note that this agreement is destroyed if one includes the finite
Higgs-boson width in the way discussed in \refse{se:poleexpansion} without 
expanding also the SMEs or appropriately adjusting the function $f(s)$
in \refeq{modborn}.

Following the treatment of \refse{se:BFMdyson}, the
matrix element for $\xxxx$ 
including $\Oa$ corrections is given similarly to \refeq{MBFM} by
\beq \label{MBFMxxxx}
\M^\xxxx = 
\left(\M^\xxxx_{\onePR} 
- 3\frac{e^2\MH^2}{4\sw^2\cw^2\MZ^2}
+ \M^\xxxx_{\boxrc}\right) \left(R_Z\right)^2
\eeq
with
\def\coupx{\frac{e^2\MH^4}{4\cw^2\sw^2\MZ^2}}
\beq \label {M1PRxxxx}
\M^\xxxx_{\onePR} = -\coupx \sum_{r=s,t,u}
\frac{(1+ F^{\chi\chi H}(r))(1+ F^{\chi\chi H}(r))}{r-\MH^2+\Si^H(r)} .
\eeq
Because $\M^\xxxx_{\onePR}$ vanishes for $s\gg\MH^2$ as $\MH^2/s$,
the boxes dominate the matrix element for $\xxxx$ at high energies.
The matrix elements for $\xxxx$ are calculated in exactly the same way as for 
$\PZ\PZ\to\PZ\PZ$. 
In contrast to $\zzzz$ the one-loop corrections to $\xxxx$ possess a
transparent form; they are explicitly presented in Appendix~\ref{se:ETexpl}.
We recall that the usage of the BFM is crucial for the ET to work for
Dyson-summed amplitudes.
Because the would-be Goldstone bosons are scalars,
no polarization vectors and no SMEs
occur,
and no unitarity cancellations between 
individual contributions take place. This simplifies the calculation
considerably.

The Higgs-boson resonance can be treated exactly as for $\zzzz$.
Since the wave functions are trivial 
constants,
and thus no split into SMEs and
invariant functions is necessary,
the ambiguity in applying the pole expansion is absent.

In order to improve the accuracy of a calculation via the ET one can
combine the full lowest-order matrix element with the $\Oa$ corrections
from $\xxxx$ resulting in
\beq\label{Mmixed}
\M_{\mixed} = \M_\Born^{\zzzzL} + \de\M_{\oneloop}^\xxxx.
\eeq
This treatment is, however, not possible if one uses Dyson summation, 
because in this case the lowest-order matrix element cannot be 
linearly separated 
from the one-loop corrections.

\subsection{Heavy-Higgs-boson effects}
\label{se:heavyHiggs}

In the literature an approximation for the matrix element
$\M^\xxxx$ by the leading contributions for $s,\MH^2\gg\MZ^2$ was 
frequently used
\cite{Pa85,Da89,Ma89,ETVVVV,Gu93,VVVVcorr}.
In this approximation the building blocks of \refeqs{MBFMxxxx} and 
\refeqf {M1PRxxxx} take a particularly simple form.
In the BFM we find in this limit
\beqar
\label{eq:HHxxxx1}
\Si^H(r) &=&
\frac{\al}{4\pi}\frac{\MH^2}{8\sw^2\MW^2}\Bigl\{
3\MH^2\big[3 B_0(r,\MH,\MH) + B_0(r,0,0) \nl
&&{} - 3 B_0(\MH^2,\MH,\MH) - \Re\{B_0(\MH^2,0,0)\}\big] 
\Bigr\}
+ \de Z_{\hat H}(r-\MH^2) 
, \nl
 F^{\chi\chi H}(r) &=& -\frac{\al}{4\pi}\frac{\MH^2}{8\sw^2\MW^2}\Bigl\{
2\MH^2\big[3C_0(r,0,0,\MH,\MH,0) + C_0(r,0,0,0,0,\MH)\big] \nl
&&{}+ 3 B_0(r,\MH,\MH) + 5 B_0(r,0,0) + 4 B_0(0,0,\MH) \nl
&&{} - 9 B_0(\MH^2,\MH,\MH) - 3\Re\{B_0(\MH^2,0,0)\} 
- \frac{1}{2}\Bigr\}
+ \frac{1}{2}\de Z_{\hat H} + \de Z_{\hat\chi}
, \nl
 \M^{\xxxx}_{\boxrc}&=&
\alpha^2 \frac{\MH^4}{32\sw^4\MW^4}\Bigl\{
2\MH^4\big[D_0(0,0,0,0,s,t,\MH,0,\MH,0) \nl
&&{}+ D_0(0,0,0,0,s,t,0,\MH,0,\MH)\big] \nl
&&{}+4\MH^2\big[C_0(s,0,0,\MH,\MH,0) + 3 C_0(s,0,0,0,0,\MH)\big] \nl
&&{}+  B_0(s,\MH,\MH) + 11 B_0(s,0,0)  \nl
&&{} - 9 B_0(\MH^2,\MH,\MH) - 3\Re\{B_0(\MH^2,0,0)\} - 1\nl
&&{} +\; (s\rightarrow t,t\rightarrow u) \;+\;
           (s\rightarrow u,t\rightarrow s)\Bigr\} 
     + 2\de Z_{\hat\chi}, \nl
R_Z &=& 1,
\eeqar
where $B_0$, $C_0$, and $D_0$ are scalar one-loop functions
\cite{adhab,tH79}.
The wave-function renormalization constants read in the BFM
\beq
\de Z_{\hat H} = \de Z_{\hat\chi} =
-\frac{\al}{4\pi}\frac{\MH^2}{8\sw^2\MW^2}.
\eeq
The above results are in agreement with those of \citere{ETVVVV}.
Since the approximation \refeq{eq:HHxxxx1} 
merely involves corrections of \OaMH, 
it follows from power counting 
(see \citere{Gr95} and references therein) 
that only diagrams with 
internal scalar lines contribute.
We note in passing that the terms of \OaMH\ originate entirely from the
SU(2) sector of the SM, i.e.\ they could also be obtained from the
corresponding reaction 
$\PWO\PWO\to\PWO\PWO$ in the pure SU(2) gauge theory.

We have checked that in the limit $\MZ^2\ll s\ll\MH^2$ the amplitude
$\M^{\xxxx}$ reduces to
\beq
\label{eq:HHxxxx2}
\M^{\xxxx} = \sum_{r=s,t,u} \frac{\al^2 r^2}{16\sw^4\MW^4}
\left[ \ln\left(\frac{\MH^2}{-r-\ri\eps}\right)
      +\frac{3\sqrt{3}\pi}{2}-\frac{26}{3} \right],
\eeq
as already given in 
\citeres{Pa85,Da89}.
In this context we remark that the result \refeq{eq:HHxxxx2} can be most
easily obtained from the general structure of the heavy-Higgs-boson
limit of the SM. 
The matrix element $\M^{\xxxx}$ \refeq{eq:HHxxxx2} 
gets contributions only 
from ${\cal L}_{4}$ and ${\cal L}_{5}$ of the effective Lagrangian of
\citere{HHeff}, which quantifies the heavy Higgs-boson
effects, and from the
(three) irreducible graphs in the gauged non-linear $\si$-model which
contain only quartic scalar couplings.

\section{Discussion of numerical results}
\label{se:numres}

\subsection{Computational details}

For the calculations we use the following parameter set \cite{pdg}:
\begin{equation}
\arraycolsep 8pt    
\begin{array}[b]{lll}
\alpha ^{-1}=137.0359895\,, &
\MZ=91.188\GeV\,, &
\MW=80.26\GeV\,, \\[4pt]
\Me=0.51099906\MeV\,, &
\Mu=47.0\MeV\,, &
\Md=47.0\MeV\,, \\
m_\mu =105.658389\MeV\,, &
\Mc=1.55\GeV\,, &
\Ms=150\MeV\,, \\  
m_\tau =1771.1\MeV\,, &
\Mt=180\GeV\,, &
\Mb=4.5\GeV\,. \\
\end{array}
\end{equation}
The masses of the light quarks 
are adjusted such that the 
experimentally measured hadronic vacuum polarization is reproduced
\cite{Ei95}.
For a Higgs-boson with a mass of $\MH=700\GeV$ these parameters yield the
lowest-order decay width 
$\Gamma_\PH\approx175.29\GeV$, \ie about one
fourth of the mass.

The various independent calculations described in \refse{se:calframe}
agree numerically typically to
$\sim 10$ digits apart from the regions close to the boundaries of phase
space.
At these boundaries the 
reduction of tensor integrals to scalar integrals breaks down.
We avoid these regions by using the angular cut
$\theta_{\cut}=10^\circ$, which also removes the Landau 
singularities in the fermionic boxes for energies above about $500\GeV$.

\subsection{Corrected \css}

\begin{figure}
\epsfig{figure=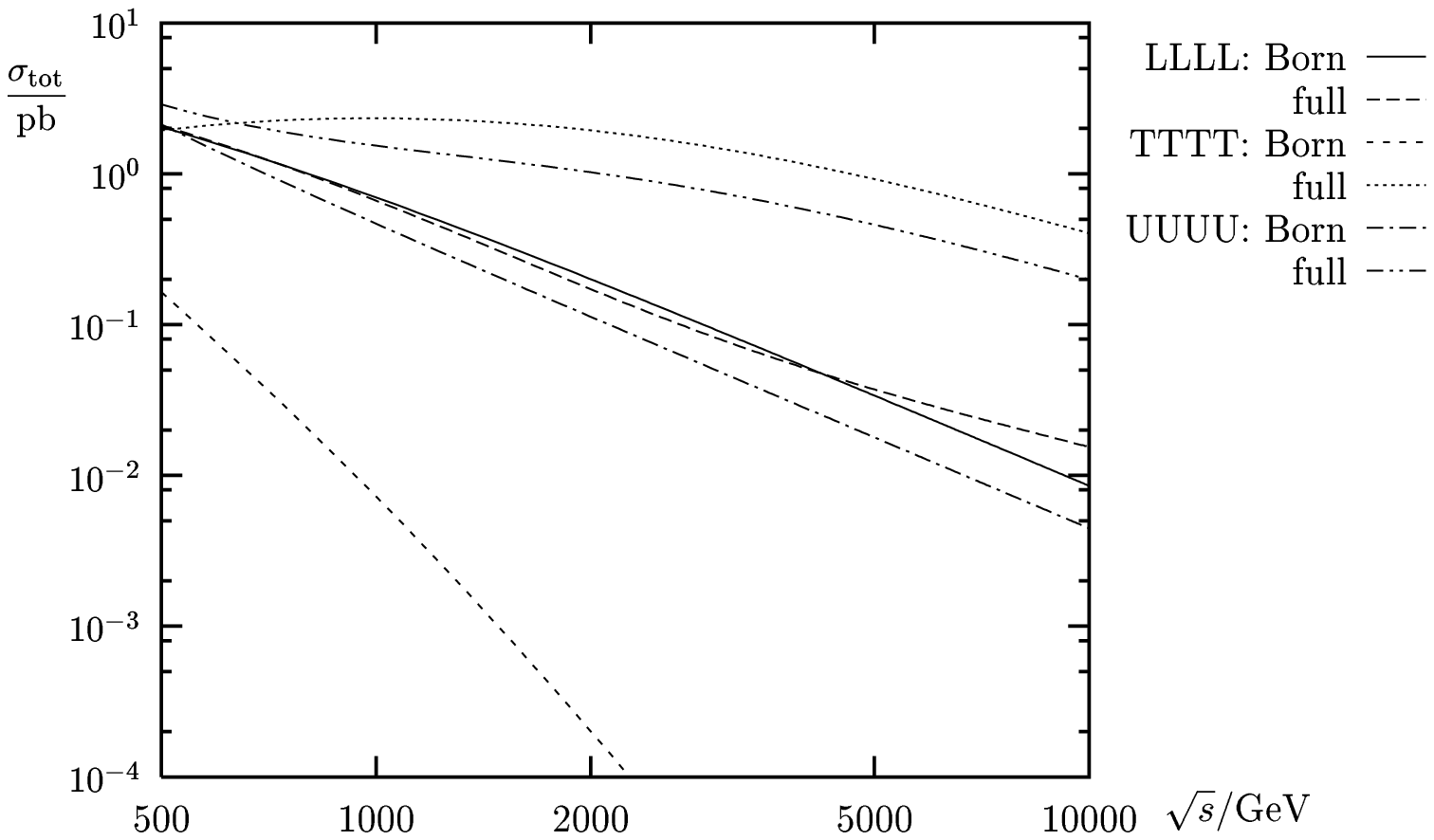}
\caption{Lowest-order and corrected integrated \css\ for various polarizations  
at $\MH=100\GeV$}
\label{fi:tot.MH=100}
\vspace{2em}
\epsfig{figure=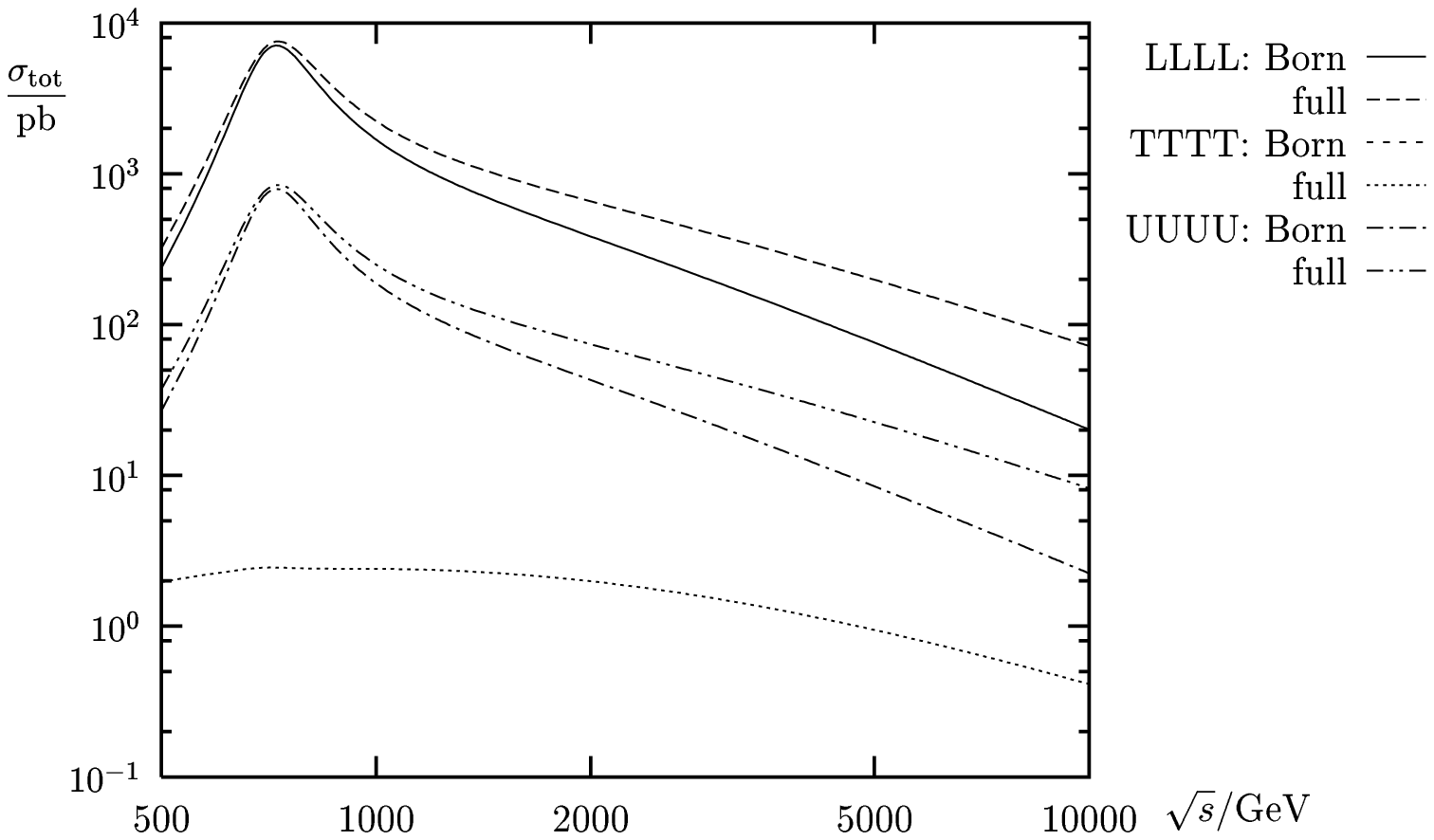}
\caption{Lowest-order and corrected integrated \css\ for various polarizations  
at $\MH=700\GeV$ (The lowest-order TTTT \cs\ is not visible.)}
\label{fi:tot.MH=700}
\end{figure}

The integrated
\css\ for unpolarized, purely transverse, and purely longitudinal
\PZ~bosons in lowest order and including the one-loop corrections
are shown in \reffis{fi:tot.MH=100} and \ref{fi:tot.MH=700} for $\MH=100\GeV$
and $\MH=700\GeV$, respectively (repeating information from 
\reffis{fi:born.MH=100}, \ref{fi:born.MH=700}, \ref{fi:onel.MH=100}, 
and \ref{fi:onel.MH=700}).
In the case of $\MH=100\GeV$ no finite Higgs-boson width is introduced;
for $\MH=700\GeV$ we apply Dyson summation within the BFM using the
renormalization scheme of \citere{bgflong}.
The Higgs-boson-mass dependence of the \cs\ for purely 
transverse \PZ~bosons is below
10\% including the Higgs-boson-resonance effects.
The corresponding lowest-order
\cs\ is very small at high energies and not 
visible in \reffi{fi:tot.MH=700}.

\begin{figure}
\epsfig{figure=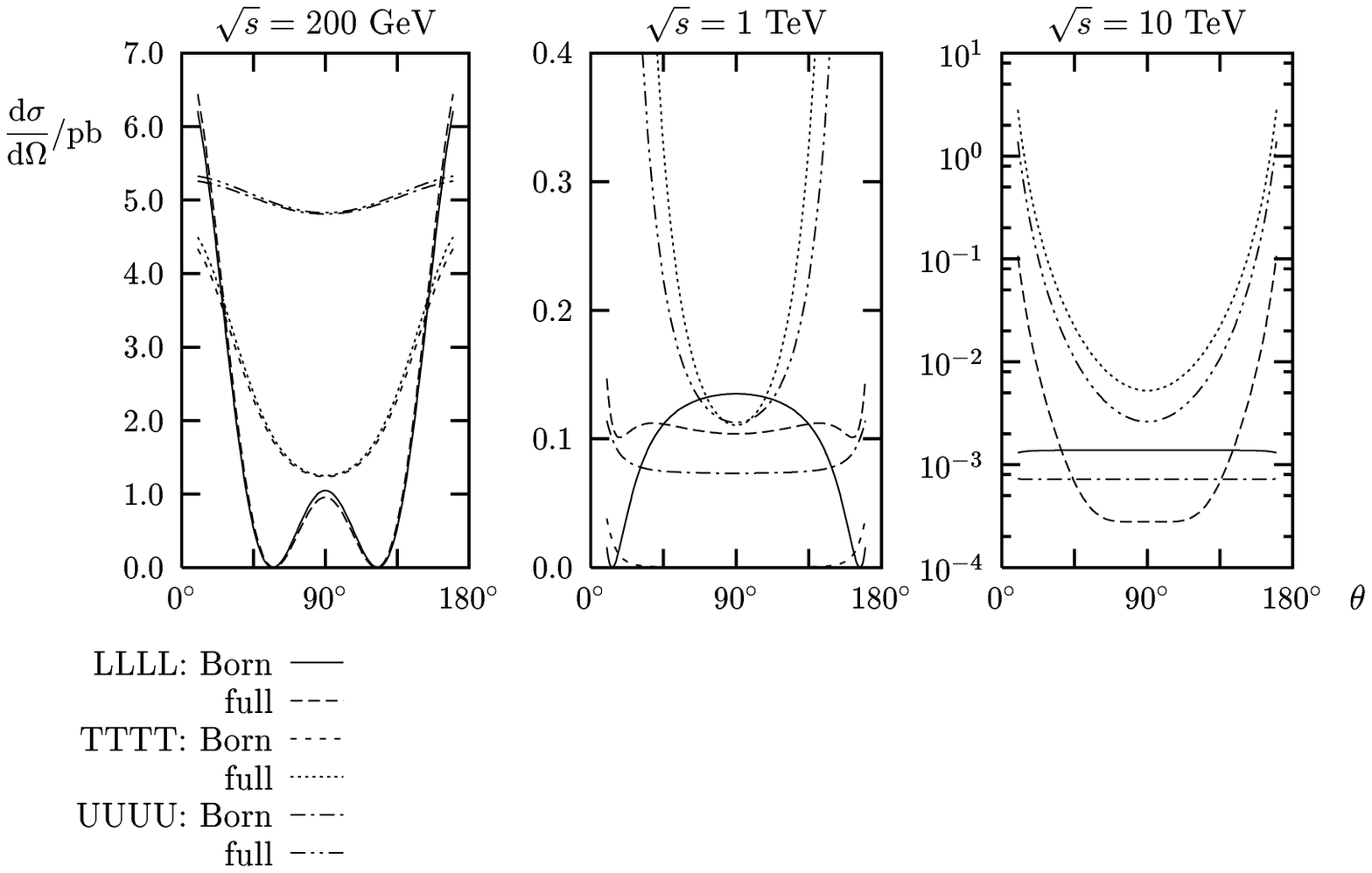, width=16cm}
\caption{Lowest-order and corrected differential \css\ for various 
polarizations and CMS energies at $\MH=100\GeV$}
\label{fi:diff.MH=100}
\vspace{2em}
\epsfig{figure=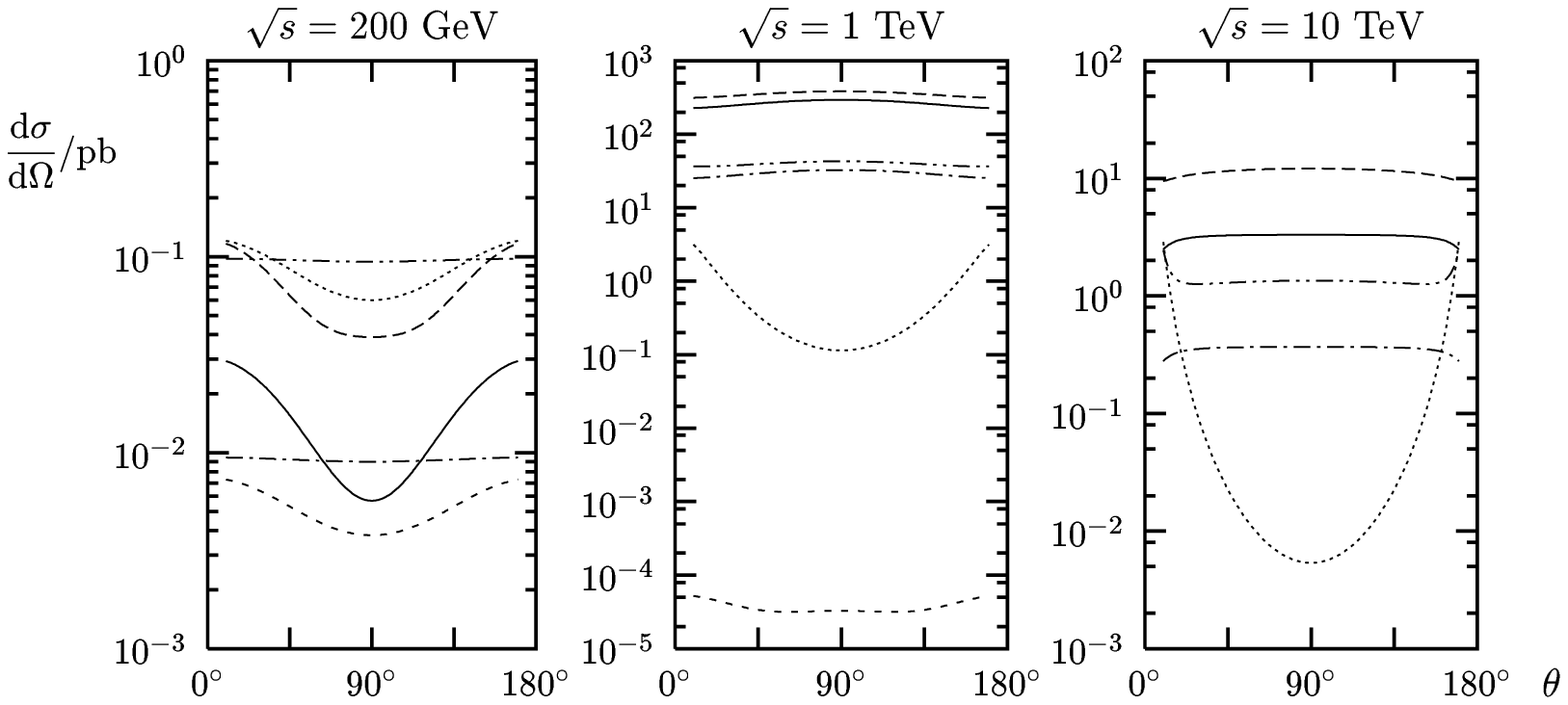, width=16cm}
\caption{Lowest-order and corrected differential \css\ for various 
polarizations (as indicated in \reffi{fi:diff.MH=100})
and CMS energies at $\MH=700\GeV$}
\label{fi:diff.MH=700}
\end{figure}

The differential \css\ for various energies are shown in
\reffis{fi:diff.MH=100} and \ref{fi:diff.MH=700}.
For small energies the angular variation of the \css\ stays within one
order of magnitude.
For high energies and a small Higgs-boson mass
the corrected \css\ are strongly peaked in
the forward and backward directions, 
while the lowest-order \css\
are relatively flat in the considered angular region. 
The \cs\ for purely longitudinal gauge bosons has kinematical zeros 
if $\MH\lsim \sqrt{1+\smash{\sqrt3}\rule{0mm}{3.8mm}}\MZ \approx 150\GeV$,
which move towards the forward and backward directions with increasing energy.
For a large Higgs-boson mass the \cs\ for purely longitudinal \PZ~bosons,
which dominates in this regime, becomes flat and therefore also the
unpolarized \cs.

\subsection{Higgs-boson resonance}

\begin{figure}
\epsfig{figure=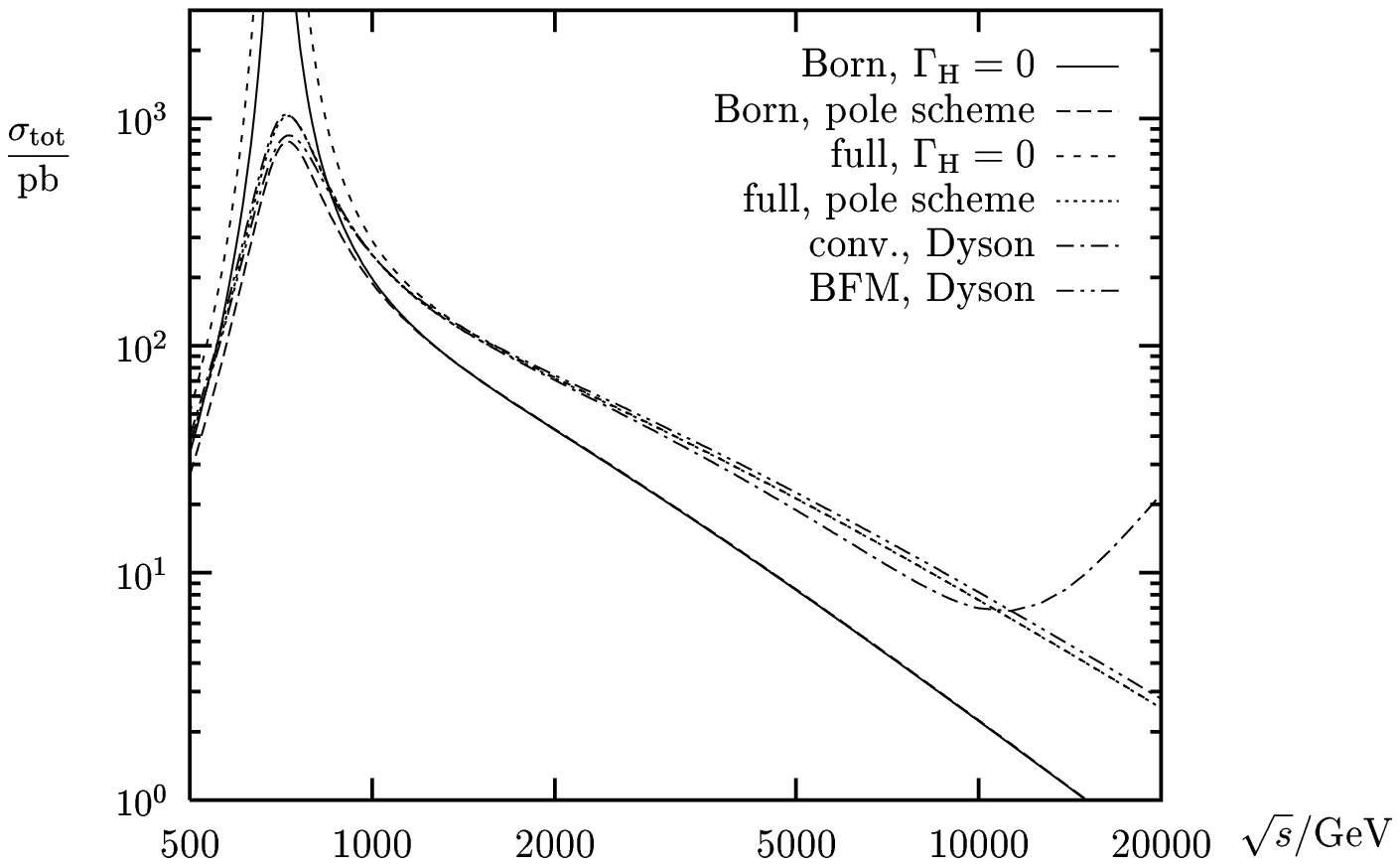}
\caption{Integrated unpolarized \cs\ 
at $\MH=700\GeV$ for various treatments of the Higgs-boson resonance}
\label{fi:resotreatments}
\vspace{2em}
\epsfig{figure=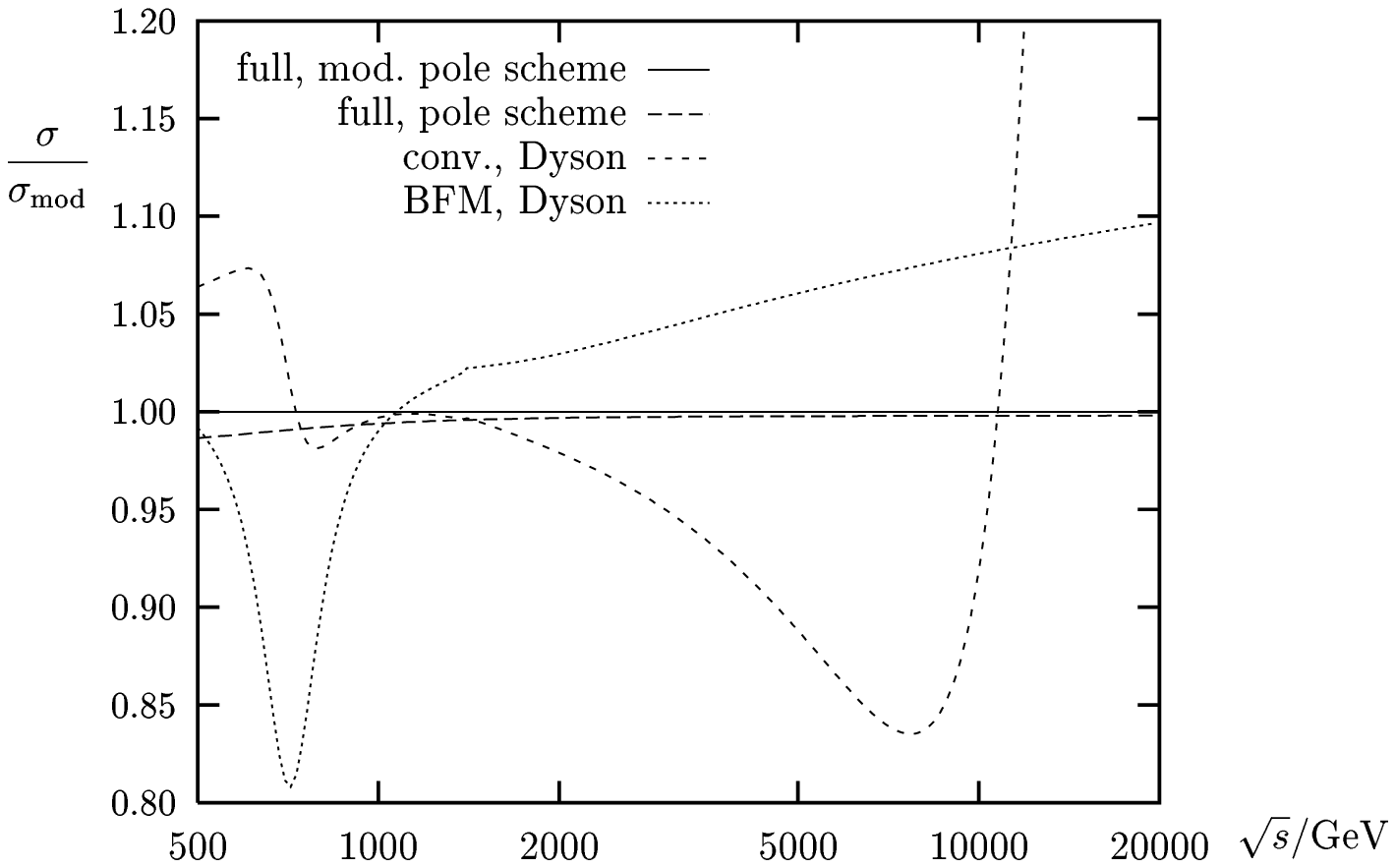}
\caption{Relative deviation of various treatments of the Higgs-boson resonance
from the modified pole-scheme result for the 
integrated unpolarized \cs\ at $\MH=700\GeV$}
\label{fi:resotreatments.relative}
\end{figure}

In \reffis{fi:resotreatments} and \ref{fi:resotreatments.relative}
we compare
several different treatments of the Higgs-boson resonance using 
$\MH=700\GeV$. 
We include the lowest-order (Born, $\Ga_\PH=0$) and the corrected (full,
$\Ga_\PH=0$) \css\ for vanishing Higgs-boson width for reference. 
We show the lowest-order (Born, pole scheme) and the 
corrected (full, pole scheme) \cs\ in the pole-scheme treatment given 
in \refeqs{resonancetreatment1} and
\refeqf{resonancetreatment2}
as well as the corrected \cs\ in the modified pole
scheme (full, mod.\ pole scheme) 
according to \refeqs{resonancetreatment3} 
with $f(s)=\MH^2/s$. In addition we give
the \css\ resulting from Dyson summation according to 
\refeq{MDyson} and
\refeq{MBFM}
in the BFM (BFM, Dyson) 
and the corresponding one in the conventional formalism (conv., Dyson). 
Apart from the Dyson-summed \css\ all others are
identical in the BFM and in the conventional formalism.
Since the unpolarized \cs\ is dominated by the one for purely
longitudinal \PZ~bosons for a large Higgs-boson mass, 
\reffi{fi:resotreatments} 
holds essentially also for the latter 
\cs\ after multiplying by a factor nine.

The crucial differences between the various treatments can already be
seen in \reffi{fi:resotreatments}, which shows the integrated cross-section.
Owing to the crude resolution the pole-scheme \css\ with or without 
$\Oa$ corrections cannot be separated
from the corresponding \css\ for $\Gamma_\PH=0$ at high energies.
The deviation of the Dyson-summed BFM \cs\ is due to higher-order corrections 
that become increasingly important with energy. 
The Dyson-summed conventional \cs\ deviates more for energies above a
few TeV and becomes completely wrong for energies higher than $10\TeV$.
This results from the violation of the Ward identities which
leads to unitarity violation at high energies.

The differences between the various treatments
of the Higgs-boson 
resonance can be seen more clearly 
in \reffi{fi:resotreatments.relative},
where the corrected \css\ are shown normalized to the one in the
modified pole scheme. The difference between the pole scheme and the
modified pole scheme is below 2\% and 
becomes small at high energies. 
Note, however, that by using dimensionful SMEs the pole-scheme \css\
could become completely wrong at high energies owing to spurious
unitarity-violating terms.

In the resonance region 
the Dyson-summed \css\ deviate from the \css\ in the  
modified pole scheme by up to 19\% and 7\% in the BFM 
and the conventional formalism, respectively.
This difference is due to the fact that our calculation near the
resonance is only of $\O(1)$ accuracy since the lowest-order
contribution in the resonance denominator vanishes on resonance
(\cf \refse{se:poleexpansion}).
The size of these differences and the correction of 24\% of the pole
scheme calculation on resonance set the typical scale for the missing $\Oa$
corrections in the resonance region.

\subsection{The \cs\ for \boldmath{$\zzzzL$} and the equivalence theorem}

\begin{figure}
\epsfig{figure=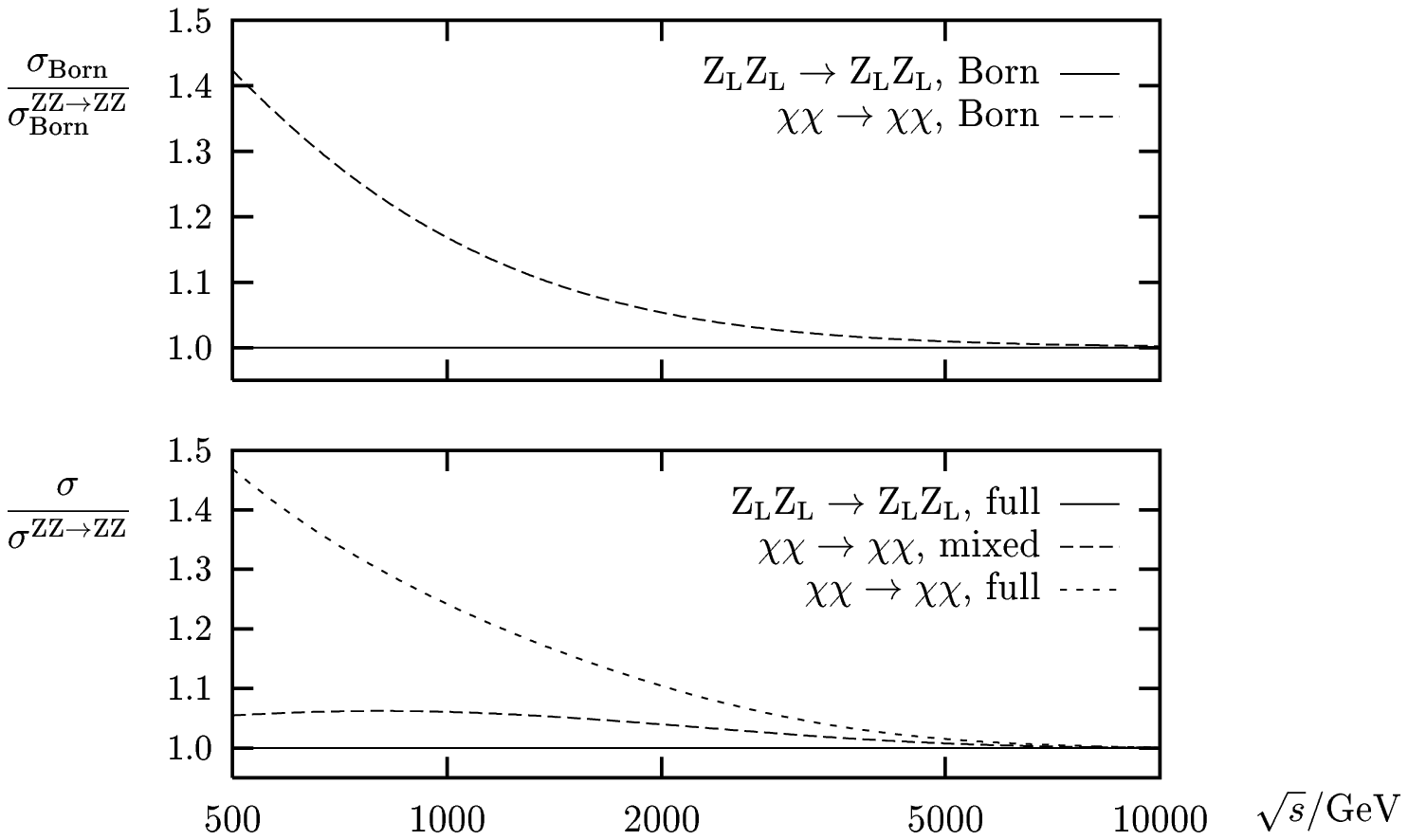}
\caption{Relative deviations of the ET predictions for $\zzzzL$ 
at $\MH=100\GeV$}
\label{fi:et.MH=100}
\vspace{2em}
\epsfig{figure=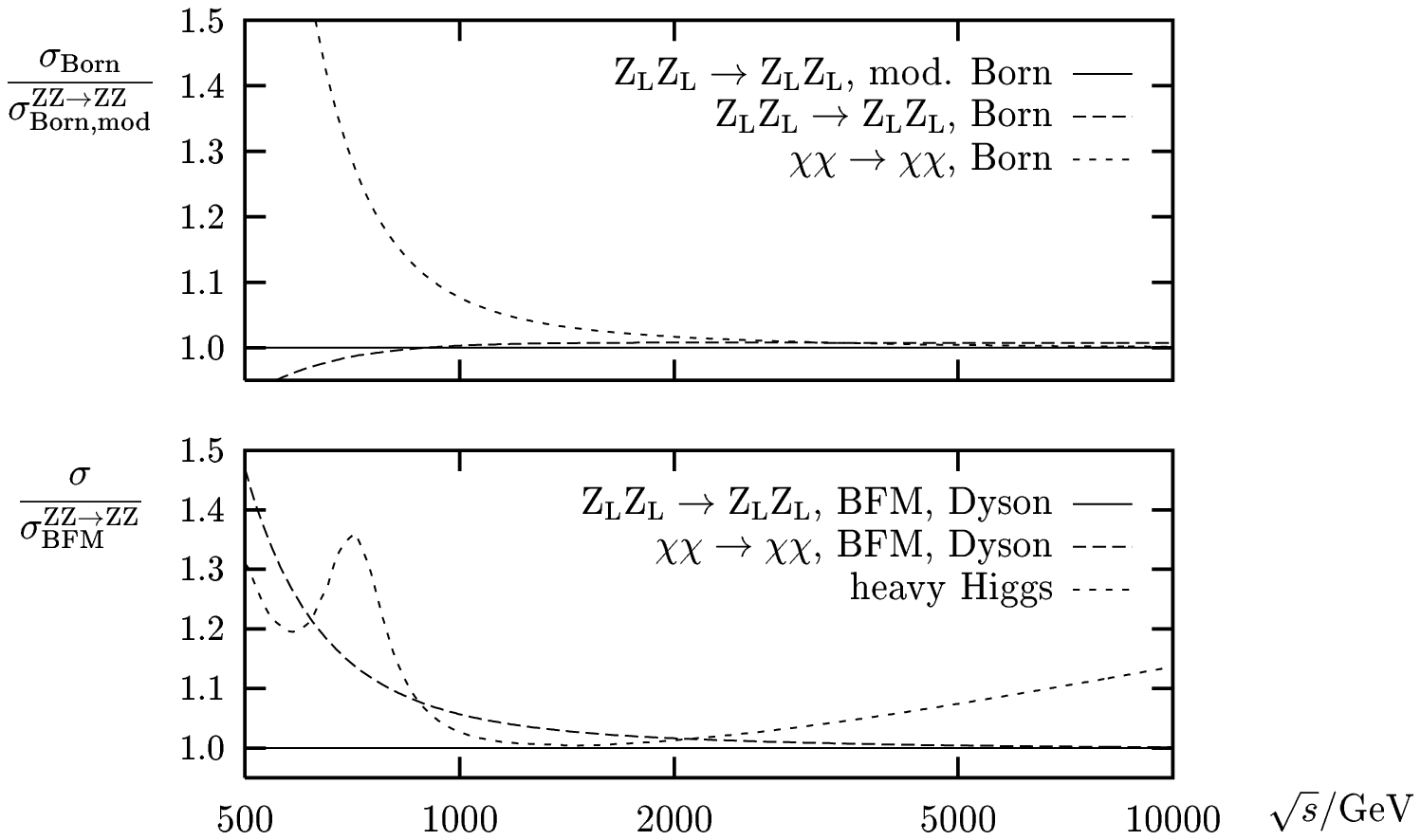}
\caption{Relative deviations of the ET predictions for $\zzzzL$ 
at $\MH=700\GeV$}
\label{fi:et.MH=700}
\end{figure}

Finally,
we want to investigate the numerical accuracy of the ET.
We distinguish the cases without and with a Higgs-boson resonance.
In \reffi{fi:et.MH=100} we consider the case of no Higgs-boson resonance
($\MH=100\GeV$). 
We show the lowest-order \cs\ (Born) calculated from the 
ET normalized to the lowest-order \cs\ for $\zzzzL$, the fully corrected
\cs\ (full) calculated from the ET 
and the \cs\ obtained from the matrix element \refeq{Mmixed} (mixed)
both normalized to the fully corrected
\cs\ for $\zzzzL$. 
The quality of the ET at $E_\CMS=1\TeV\ (2\TeV)$ is about 17\% (5\%) 
for the lowest order, 24\% ($10\%$) for one-loop,
and 6\% (4\%) for one-loop mixed.
As expected, the  one-loop mixed approximation is substantially better
than the simple ET \cs.

In \reffi{fi:et.MH=700} we investigate the accuracy of the ET in the 
presence of a Higgs-boson resonance at $\MH=700\GeV$. 
We show again the ratios of the lowest-order and
corrected \css\ obtained  using the ET and from the direct calculation. 
In the lowest-order \cs\ we include the finite width 
{\naive}ly (Born) and in the
modified pole scheme (mod.\ Born) \refeq{modborn}.
The lowest-order \cs\ from the ET approaches that of the modified pole
scheme at high energies. Including the finite width {\naive}ly
leads to a \cs\ that deviates at high energies from these two \css\ by a
factor $1+\Ga_\PH^2/9\MH^2\approx 1.007$ for $\MH=700\GeV$
(\cf \refse{se:poleexpansion}).
For the corrected \css\ we have applied Dyson summation within the BFM.
Because of the Dyson summation the mixed case does not make
sense anymore. Instead we show the \cs\ resulting from the \OaMH\
approximation of the RCs
\refeq{eq:HHxxxx1} normalized to the fully corrected \cs\ for $\zzzzL$.
The ET works much better for a heavy Higgs boson.
At $E_\CMS=1\TeV\ (2\TeV)$ we now find a deviation of 8\% (2\%) for the
lowest order [using the modified Born according to \refeq{modborn}]
and 6\% (2\%) for the corrected \cs. 
For energies above $2\TeV$ the deviation between the corrected \css\ is
practically equal to the deviation between the lowest-order \css.
The \OaMH\ approximation \refeq{eq:HHxxxx1} works well in the regime
$\MZ^2\ll\MH^2\ll s\ll \MH^4/\MZ^2$, where the upper limit for the energy
results from the neglect
of corrections proportional to $s/\MZ^2$ with
respect to the ones proportional to $\MH^4/\MZ^4$.
For $\MH=700\GeV$ this restricts the
energy to $\sqrt{s}\sim 1-3\TeV$,  
which is nicely reflected in the figure.

\section{Conclusions}
\label{se:concl}

Owing to the strong sensitivity to the gauge-boson and scalar
self-interactions, scattering of massive gauge bosons found continuous
interest in the literature, where the emphasis was directed to
strong-coupling effects for longitudinally polarized gauge bosons. We
have supplemented the existing results for the enhanced radiative
corrections of order \OaMH\ by the complete 
${\cal O}(\al)$ corrections to $\zzzz$ for arbitrarily polarized Z
bosons. 

At high energies the
radiative corrections are found to be large, at several TeV they are 
typically of the order of the lowest-order \css. 
Whereas the \cs\ for purely transverse 
Z bosons at high energies 
is totally negligible in lowest order, the  
corrections enhance this \cs\ such that it becomes one of the dominating
channels.

The introduction of a finite Higgs-boson width in order to 
describe the resonance well is a non-trivial task. 
We have compared different approaches, viz.\ different variants of 
the Laurent expansion about the complex pole and the Dyson summation of
self-energies, where the latter has been performed both in the
conventional 
formalism as well as in the background-field formalism. 
{}From a theoretical point of view, the
background-field approach is the most convincing one, since it
naturally guarantees a reasonable cross-section also far above the
resonance, where the validity of Ward identities is crucial to imply
the necessary gauge cancellations.
However, in order to obtain a relative precision of $\Oa$ on resonance 
one would have to perform a complete two-loop calculation.
In order to obtain the same precision on resonance in the pole scheme
only 
the imaginary part of the self-energy has to be evaluated at two loops.
However, as the pole scheme and 
the other mentioned methods do not care about the Ward
identities, theoretical uncertainties may get out of control in the presence
of gauge cancellations. 
Using the pole scheme carelessly can lead to
unitarity-violating terms at high energies, and Dyson summation
within the conventional formalism in fact yields a totally wrong 
cross-section in the high-energy limit.

We have investigated longitudinal Z-boson scattering $\zzzzL$ in more
detail and performed a complete $\Oa$ calculation using the 
Goldstone-boson equivalence theorem. 
For a center-of-mass energy of $1\TeV$ ($2\TeV$) the deviation of the 
equivalence theorem from the exact $\Oa$ result is 
about 24\% (10\%) and 6\% (2\%) for a Higgs-boson mass $\MH$ of 
$100\GeV$ and $700\GeV$, respectively, with an asymptotic approach in the
high-energy limit.
The frequently used approximation by the enhanced corrections of \OaMH\
for a heavy Higgs boson is good for energies of a few TeV but 
gets worse with increasing energy.

Although Z-boson scattering is the simplest representative of
massive gauge-boson scattering, it 
contains the typical
features such as the Higgs-boson resonance and enhanced heavy-Higgs-boson
corrections.
In contrast to other gauge-boson scattering processes, 
the lowest-order \css\ for transverse \PZ-boson scattering are
suppressed and no real photon radiation needs to be considered in
$\zzzz$.
Nevertheless, we expect that our results 
at least qualitatively
carry over to the other massive gauge-boson scattering reactions.

\def\appendix{\setcounter{section}{0} \setcounter{subsection}{0}
              \setcounter{equation}{0}
              \@addtoreset{equation}{section}
              \def\thesection{\Alph{section}}
              \def\theequation{\thesection\arabic{equation}}}

\appendix
\section*{Appendix}

\section{Discussion of the Landau singularity in box diagrams}
\label{app:d0sing}

In \refse{subse:sing} we have briefly discussed the Landau singularity
which occurs in some fermionic box diagrams if the fermion mass $m$ fulfills
$m<\MZ/2$. Although we have argued that this singularity is
unphysical
and only caused by the use of  
the equivalent vector-boson approximation for pp or ee collisions, 
it is nevertheless interesting to investigate some formal
properties of the singularity. 

{}From general considerations 
(e.g.\ about unitarity)
one expects that the singularity drops out 
in the fully inclusive \cs, \ie
if all possible final states are taken into account. We 
have verified this
compensation by explicitly calculating the singular contributions of
\reffi{fig:diagsing} to the inclusive cross-section $\PZ\PZ\to 4\Pf$.
\unitlength 1cm
\begin{figure}
\begin{center}
\begin{picture}(13.5,10)
\put(-4.0,-15.2){\includegraphics{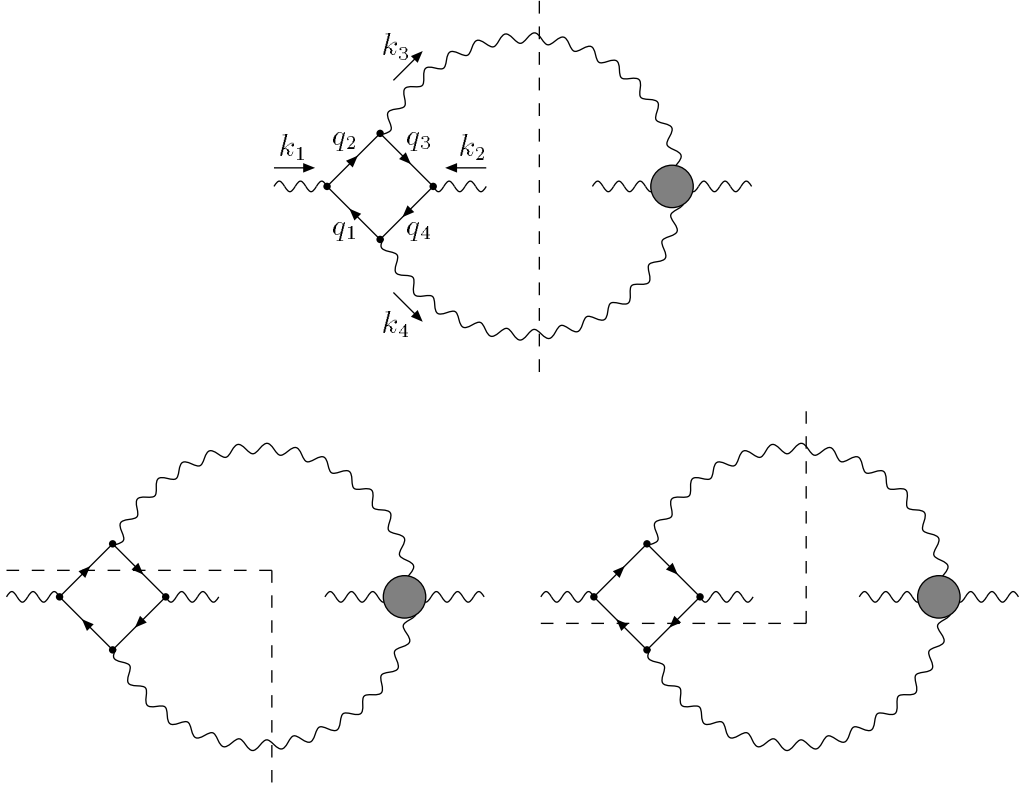}}
\end{picture}
\end{center}
\caption{Generic form of some singular contributions to 
\protect{$\PZ\PZ\to\PZ\Pf\Pfbar$}, which are related by cutting rules.}
\label{fig:diagsing}
\end{figure}
The shaded circle in \reffi{fig:diagsing} represents
any regular graph for \protect{$\PZ\PZ\to\PZ\PZ$},
i.e.\ only the cuts that are explicitly shown in \reffi{fig:diagsing}
are relevant for the singular contributions.
If the shaded circle also contains the singularity 
more cuts have to be considered.
We restrict ourselves to the case where the produced Z bosons 
are on their mass shell, $k_3^2=k_4^2=\MZ^2$. 

The singularity in the loop integral of the upper graph of
\reffi{fig:diagsing} stems from integration momenta $q_1\sim\hat q$ with
\beq
\hat q^\mu = -\frac{1}{2}(k_1-k_3)^\mu
+\frac{t-2\MZ^2}{2s}(k_3+k_4)^\mu, \qquad
\hat q^2 = \frac{4\MZ^4-tu}{4s},
\eeq
and occurs for $\hat q^2\to m^2$, which follows from the Landau
equations for the integral. Thus, the singular contribution
of the virtual graph is simply obtained by setting $q_1\to\hat q$ in the
numerator such that the remaining integral is proportional to the
$D_0$ function, the singular contribution of which 
is given in \refeq{landausing}.

The singular contributions in the lower graphs of \reffi{fig:diagsing}
occur in the phase-space integrals of the produced fermion-antifermion
pair and can be obtained from the corresponding scalar integrals
analogously to the loop integral. For instance the relevant scalar
integral for the lower left graph is given by
\beq
I = 
\int\frac{\mathrm{d}^3{\bf k}}{(2\pi)^3 2k^0}\,
\int\frac{\mathrm{d}^3{\bf k}'}{(2\pi)^3 2k^{\prime 0}}\,
(2\pi)^4\de^{(4)}(k_3-k-k')\,
\frac{1}{(q_1^2-m^2+\ri\eps)(q_4^2-m^2+\ri\eps)},
\eeq
where
\beq
q_1 = k-k_1, \qquad
q_4 = k-k_1+k_4 = k_2-k', \qquad
k^2=k^{\prime 2}=m^2.
\eeq
Inspection of the Landau equations for $I$ reveals that the singularity in 
$I$ originates from 
a point in phase space where the 
fermion momenta ${\bf k}$ and ${\bf k}'$ are 
coplanar with the scattering plane spanned by the ${\bf k}_l$.
Explicit calculation yields
\beq
\left.I\right|_{\mathrm{sing}} = 
-\frac{\ri}{16\pi^2}\left.D_0\right|_{\mathrm{sing}},
\eeq
which is also valid for the corresponding scalar integral for the lower
right graph of \reffi{fig:diagsing}.

The relation between the singular contributions of $D_0$ and $I$ guarantees
the cancellation of the singularity in the sum of the graphs of
\reffi{fig:diagsing} if the produced Z bosons are on shell. 
However, 
the cancellation in general is incomplete if
individual flavors 
or spins in the fermionic final state are observed or 
if phase-space cuts are applied. 

Therefore, we conclude that a careful
analysis of the actual physical realization of the underlying process is
mandatory if such singularities appear for physical situations. 
For the subprocess $\zzzz$ this means that one has 
to go one step back and to consider the full reaction 
including the production mechanism of the incoming \PZ~bosons
in more detail.

\section{One-loop corrections to \boldmath{$\xxxx$}}
\label{se:ETexpl}

In \refse{se:heavyHiggs} we have explicitly given the leading
corrections to longitudinal Z-boson scattering 
in the limit $s,\MH^2\gg\MZ^2$. 
More generally, the complete 
corrections to $\xxxx$ also take a 
relatively simple form in contrast to the formulae for $\zzzzL$,
\beqar
\label{eq:xxxxb}
\Si^H(r) &=&
\frac{\al}{4\pi}\frac{1}{8\sw^2\MW^2}\Bigl\{
9\MH^4 B_0(r,\MH,\MH) \nl
&& {}
+ 2(\MH^4+4\MH^2\MW^2+12\MW^4-8r\MW^2) B_0(r,\MW,\MW) \nl
&& {}
+ (\MH^4+4\MH^2\MZ^2+12\MZ^4-8r\MZ^2) B_0(r,\MZ,\MZ) \nl
&& {}
+ 3\MH^4 B_0(0,0,\MH) 
+ 2\MW^2(\MH^2+6\MW^2) B_0(0,0,\MW) \nl
&& {}
+ \MZ^2(\MH^2+6\MZ^2) B_0(0,0,\MZ) 
-24\MW^4-12\MZ^4
\nl && {}
+\sum_f 4N^{\mathrm{c}}_f m_f^2\Bigl[
(r-4m_f^2) B_0(r,\Mf,\Mf)  - 2 \Mf^2 B_0(0,0,\Mf)
\Bigr] \, \Bigr\} \nl
&& {}
-\de\MH^2 + \de Z_{\hat H}(r-\MH^2), \nn\\[.5em]
F^{\chi\chi H}(r) &=& -\frac{\al}{4\pi}\frac{1}{8\sw^2\MH^2\MW^2}\Bigl\{ \nl
&& \phantom{ {}+{} }
6\MH^2\left[(\MH^2-\MZ^2)^2+2\MZ^2(r-2\MZ^2)\right]
C_0(r,\MZ^2,\MZ^2,\MH,\MH,\MZ) \nl
&& {}
+ 2(\MH^6-3\MH^2\MZ^4-6\MZ^6+4r\MH^2\MZ^2)
C_0(r,\MZ^2,\MZ^2,\MZ,\MZ,\MH) \nl
&& {}
+ 8\MW^2(\MH^2+4\MW^2)(r-2\MZ^2)
C_0(r,\MZ^2,\MZ^2,\MW,\MW,\MW) \nl
&&{}
+ 3\MH^2(\MH^2+2\MZ^2)B_0(r,\MH,\MH) \nl
&&{}
+ 2(\MH^4+4\MH^2\MW^2+12\MW^4)B_0(r,\MW,\MW) \nl
&&{}
+ (3\MH^4+6\MH^2\MZ^2+8\MZ^4)B_0(r,\MZ,\MZ) \nl
&&{}
+ 4(\MH^2-\MZ^2)^2B_0(\MZ^2,\MH,\MZ) 
-16\MW^4-8\MZ^4
\nl &&{}
+ \sum_f 8N^{\mathrm{c}}_f \Mf^4 
\Bigl[(2\MZ^2-r)C_0(r,\MZ^2,\MZ^2,\Mf,\Mf,\Mf) - 2 B_0(r,\Mf,\Mf) \Bigr] 
\, \Bigr\}
\nl &&{}
+ \de Z_e - \frac{\de\sw^2}{2\sw^2} - \frac{\de\MW^2}{2\MW^2}
+ \frac{\de\MH^2}{\MH^2} + \frac{e}{2\sw}\frac{\de t}{\MW\MH^2}
+ \frac{1}{2}\de Z_{\hat H} + \de Z_{\hat\chi}, \nn\\[.5em]
\M^{\xxxx}_{\boxrc}&=&
\alpha^2 \frac{1}{32\sw^4\MW^4}\Bigl\{
2\left[(\MH^2-\MZ^2)^2+2\MZ^2(s-2\MZ^2)\right]^2 \nl
&&\phantom{ {}+{} }
\times D_0(\MZ^2,\MZ^2,\MZ^2,\MZ^2,s,t,\MH,\MZ,\MH,\MZ) \nl
&&{} 
+2\left[(\MH^2-\MZ^2)^2+2\MZ^2(t-2\MZ^2)\right]^2 \nl
&&\phantom{ {}+{} }
\times D_0(\MZ^2,\MZ^2,\MZ^2,\MZ^2,t,s,\MH,\MZ,\MH,\MZ) \nl
&&{} 
+16\MW^4\left[(s-2\MZ^2)^2+(t-2\MZ^2)^2\right] \nl
&&\phantom{ {}+{} }
\times D_0(\MZ^2,\MZ^2,\MZ^2,\MZ^2,s,t,\MW,\MW,\MW,\MW) \nl
&&{}
+4(\MH^2+2\MZ^2)\left[(\MH^2-\MZ^2)^2+2\MZ^2(s-2\MZ^2)\right] \nl
&&\phantom{ {}+{} }
\times C_0(s,\MZ^2,\MZ^2,\MH,\MH,\MZ) \nl
&&{}
+4\left[3\MH^2(\MH^2-\MZ^2)^2+4\MZ^4(s-2\MZ^2)\right]
C_0(s,\MZ^2,\MZ^2,\MZ,\MZ,\MH) \nl
&&{}
+16\MW^2(\MH^2+4\MW^2)(s-2\MZ^2)
C_0(s,\MZ^2,\MZ^2,\MW,\MW,\MW) \nl
&&{}
+(\MH^2+2\MZ^2)^2B_0(s,\MH,\MH) \nl
&&{}
+2(\MH^4+4\MH^2\MW^2+12\MW^4)B_0(s,\MW,\MW)  \nl
&&{}
+(9\MH^4+8\MZ^4)B_0(s,\MZ,\MZ) 
-16\MW^4-8\MZ^4 \nl
&&{}
+ \sum_f 4N^{\mathrm{c}}_f \Mf^4 \Bigl[
(st-2\MZ^4) D_0(\MZ^2,\MZ^2,\MZ^2,\MZ^2,s,t,\Mf,\Mf,\Mf,\Mf) \nl
&&\phantom{ {}+{} }
+ 4(2\MZ^2-s) C_0(s,\MZ^2,\MZ^2,\Mf,\Mf,\Mf) 
     -4B_0(s,\Mf,\Mf) \Bigr] \nl
&&{} \;+\; (s\rightarrow t,t\rightarrow u) \;+\;
           (s\rightarrow u,t\rightarrow s) \; \Bigr\} \nl
&&{}
+ 2\de Z_e - \frac{\de\sw^2}{\sw^2} - \frac{\de\MW^2}{\MW^2}
+ \frac{\de\MH^2}{\MH^2} + \frac{e}{2\sw}\frac{\de t}{\MW\MH^2}
+ 2\de Z_{\hat\chi}, 
\eeqar
where the sum over $f$ extends over all fermion flavours, and
$N^{\mathrm{c}}_f$ denotes the color factor for the fermion $f$.
The scalar four-point function is defined as ($p_1+p_2+p_3+p_4=0$)
\beqar
\lefteqn{
D_0(p_1^2,p_2^2,p_3^2,p_4^2,(p_1+p_2)^2,(p_2+p_3)^2,m_1,m_2,m_3,m_4) =
}\qquad\\
&&\int\frac{\rd^{4}\!q}{\ri\pi^{2}}
\frac{1}{(q^2-m_1^2)[(q+p_1)^2-m_2^2][(q+p_1+p_2)^2-m_2^2][(q-p_4)^2-m_2^2]}.
\nn
\eeqar
Note that these results are derived within the BFM and include the
heavy-Higgs-boson corrections of \refeq{eq:HHxxxx1} as special case. 
For the sake of simplicity
the explicit expressions for the counterterms are left open; they are
easily calculated in the renormalization scheme of \citere{bgflong}.

\def\vol#1{{\bf #1}}
\def\mag#1{{\sl #1}}


\begin{thebibliography}{99}
\itemsep 2pt plus 2pt minus 1pt
\frenchspacing
 \newcommand{\ap}[3]{{\sl Ann.~Phys.} {\bf #1} (19#2) #3}
 \newcommand{\app}[3]{{\sl Acta~Phys.~Pol.} {\bf #1} (19#2) #3}
 \newcommand{\cmp}[3]{{\sl Commun. Math. Phys.} {\bf #1} (19#2) #3}
 \newcommand{\cpc}[3]{{\sl Comp. Phys. Commun.} {\bf #1} (19#2) #3}
 \newcommand{\fp}[3]{{\sl Fortschr. Phys.} {\bf #1} (19#2) #3}
 \newcommand{\ijmp}[3]{{\sl Int. J. Mod. Phys.} {\bf #1} (19#2) #3}
 \newcommand{\jetp}[3]{{\sl JETP} {\bf #1} (19#2) #3}
 \newcommand{\jetpl}[3]{{\sl JETP Lett.} {\bf #1} (19#2) #3}
 \newcommand{\jmp}[3]{{\sl J. Math. Phys.} {\bf #1} (19#2) #3}
 \newcommand{\mpl}[3]{{\sl Mod. Phys. Lett.} {\bf #1} (19#2) #3}
 \newcommand{\nc}[3]{{\sl Nuovo Cimento} {\bf #1} (19#2) #3}
 \newcommand{\nim}[3]{{\sl Nucl. Instr. Meth.} {\bf #1} (19#2) #3}
 \newcommand{\np}[3]{{\sl Nucl. Phys.} {\bf #1} (19#2)~#3}
 \newcommand{\pl}[3]{{\sl Phys. Lett.} {\bf #1} (19#2) #3}
 \newcommand{\pr}[3]{{\sl Phys. Rev.} {\bf #1} (19#2) #3}
 \newcommand{\prl}[3]{{\sl Phys. Rev. Lett.} {\bf #1} (19#2) #3}
 \newcommand{\ptp}[3]{{\sl Prog. Theo. Phys.} {\bf #1} (19#2) #3}
 \newcommand{\sjnp}[3]{{\sl Sov. J. Nucl. Phys.} {\bf #1} (19#2) #3}
 \newcommand{\zp}[3]{{\sl Z. Phys.} {\bf #1} (19#2) #3}
 \newcommand{\vj}[4]{{\sl #1~}{\bf #2} (19#3) #4}
 \newcommand{\ej}[3]{{\bf #1} (19#2) #3}
 \newcommand{\vjs}[2]{{\sl #1~}{\bf #2}}

\bibitem{Di73}
D.A. Dicus and V.S. Mathur, \pr{D7}{73}{3111}.

\bibitem{Le77}
B.W. Lee, C. Quigg and H.B. Thacker, \pr{D16}{77}{1519}.

\bibitem{Pa85}
G. Passarino, \pl{156B}{85}{231} and \np{B343}{90}{31};\\
M. Veltman and F. Yndurain, \np{B325}{89}{1}.

\bibitem{Da89}
S. Dawson and S. Willenbrock, \prl{62}{89}{1232} and \pr{D40}{89}{2880}.

\bibitem{Ma89}
W. Marciano, G. Valencia and S. Willenbrock, \pr{D40}{89}{1725};\\
L. Durand, J.M. Johnson and J.L. Lopez, \prl{64}{90}{1215} and
\pr{D45}{92}{3112}.

\bibitem{ETVVVV}
L. Durand, J.M. Johnson and P.N. Maher, \pr{D44}{91}{127};\\
T.N. Truong, \pl{B258}{91}{402}.

\bibitem{Gu93}
S.N. Gupta, J.M. Johnson and W.W. Repko, \pr{D48}{93}{2083}.

\bibitem{VVVVcorr}
R. Bouamrane, in {\sl Radiative Corrections: Results and Perspectives},
eds. N. Dombey and F. Boudjema (Plenum Press, New York, 1990) p.~533;\\
D.A. Dicus and W.W. Repko, \pl{B228}{89}{503} and \pr{D42}{90}{3660};\\
S. Dawson and G. Valencia, \np{B348}{91}{23}.

\bibitem{et}
J.M. Cornwall, D.N. Levin and G. Tiktopoulos, \pr{D10}{74}{1145}; \\
G.J. Gounaris, R. K\"ogerler and H. Neufeld, \pr{D34}{86}{3257}.

\bibitem{Ch85}
M.S. Chanowitz and M.K. Gaillard, \np{B261}{85}{379}.

\bibitem{Ya88}
Y.-P. Yao and C.-P. Yuan, \pr{D38}{88}{2237}; \\
J. Bagger and C. Schmidt, \pr{D41}{90}{264}; \\
H.-J. He, Y.-P. Kuang and X. Li, \prl{69}{92}{2619} and
\pr{D49}{94}{4842}.

\bibitem{Gr95}
C. Grosse-Knetter, \zp{C67}{95}{261}.

\bibitem{bgfet}
A. Denner and S. Dittmaier, \pr{D54}{96}{4499}.

\bibitem{Ve77}
M. Veltman, \app{B8}{77}{475}.

\bibitem{Ku96}
I. Kuss and H. Spiesberger, \pr{D53}{96}{6078}.

\bibitem{bgflong}
A.~Denner, S. Dittmaier and G. Weiglein, \np{B440}{95}{95}.

\bibitem{aaww}
A.~Denner, S. Dittmaier and R. Schuster, \np{B452}{95}{80}.

\bibitem{adhab}
A.\ Denner, \fp{41}{93}{307}. 

\bibitem{FA}
J.\ K\"ublbeck, M.\ B\"ohm and A.\ Denner, \cpc{60}{91}{165}; \\
H.\ Eck and J.\ K\"ublbeck, 
{\it Guide to FeynArts 1.0}, Universit\"at W\"urzburg
(1992); \\
H.\ Eck,  {\it FeynArts 2.0---A generic Feynman diagram generator},
Dissertation,
Universit\"at W\"urzburg (1995).

\bibitem{math} S.~Wolfram, {\it Mathematica---A System for Doing
Mathematics by Computer} (Addison-Wesley, Redwood City, CA, 1988).

\bibitem{Pa79}  G.\ Passarino and M.\ Veltman, \np{B160}{79}{151}.

\bibitem{tH79}
G.\ 't Hooft and M.\ Veltman, \np{B153}{79}{365}; \\
A.\ Denner, U.\ Nierste and R.\ Scharf, \np{B367}{91}{637}.

\bibitem{Form}
J.A.M. Vermaseren, {\it Symbolic Manipulation with FORM}, 
(CAN, Amsterdam, 1991).

\bibitem{FC}
R.\ Mertig, M.\ B\"ohm, A.\ Denner, \cpc{64}{91}{345}; \\
R.\ Mertig, {\it Guide to FeynCalc 1.0}, Universit\"at W\"urzburg (1992).

\bibitem{FF}
G.J.\ van Oldenborgh and J.A.M.\ Vermaseren, \zp{C46}{90}{425}.

\bibitem{be93} W.~Beenakker et al, \pl{B317}{93}{622} and \np{B410}{93}{245}.

\bibitem{d0pol}
L.D. Landau, \np{13}{59}{181};\\
R.J.\ Eden, P.V.\ Landshoff, D.I.\ Olive and J.C.\ Polkinghorne, 
{\it The analytic $S$-Matrix}, (Cambridge Univ.\ Press, 1966).

\bibitem{St91}
R.G. Stuart, \pl{B262}{91}{113}.

\bibitem{Ae93}
A. Aeppli, F. Cuypers and G.J. van Oldenborgh, \pl{B314}{93}{413};\\
A. Aeppli, G.J. van Oldenborgh and D. Wyler,  \np{B428}{94}{126}.

\bibitem{St96}
R.G. Stuart, preprint UM-TH-96-05, hep-ph/9603351.

\bibitem{HHeff}
M.J.~Herrero and E.~Ruiz Morales, \np{B437}{95}{319};\\
S.~Dittmaier and C.~Grosse-Knetter,
\pr{D52}{95}{7276} and \np{B459}{96}{497}.

\bibitem{pdg} 
Particle Data Group, \pr{D54}{96}{1}.
The most recent values can be obtained via WWW from http://pdg.lbl.gov.

\bibitem{Ei95} 
S.~Eidelman and F.~Jegerlehner, \zp{C67}{95}{585};\\
H.~Burkhardt and B.~Pietrzyk, \pl{B356}{95}{398}.

\end{thebibliography}
\end{document}